\documentstyle[aps,epsf]{revtex}
\input epsf.sty
\def\prb{Phys. Rev. B}
\def\prl{Phys. Rev. Lett.}

\def\be{\begin{equation}}
\def\ee{\end{equation}}
\def\ba{\begin{eqnarray}}
\def\ea{\end{eqnarray}}

\def\LSCO{La$_{2-x}$Sr$_x$CuO$_4$}
\def\LCO{La$_2$CuO$_4$}
\def\LSNiO{La$_{2-x}$Sr$_x$NiO$_{4+\delta}$}
\def\YBCO{YBa$_2$Cu$_3$O$_{7-\delta}$}

\def\BSCCO{Bi$_2$Sr$_2$CaCu$_2$O$_{8+\delta}$}
\def\C60{A$_x$C$_{60}$}
\def\LNSCO{La$_{1.6-x}$Nd$_{0.4}$Sr$_x$CuO$_{4}$}

\def\hty{high temperature superconductivity}
\def\hts{high temperature superconductors}
\begin{document}
\draft
\flushbottom
\twocolumn[
\hsize\textwidth\columnwidth\hsize\csname @twocolumnfalse\endcsname

\title{Spin-Gap Proximity Effect Mechanism of High Temperature 
Superconductivity} 
\author{V.~J.~Emery }
\address{
Dept. of Physics
Brookhaven National Laboratory
Upton, NY  11973}
\author{S.~A.~Kivelson and O.~Zachar}
\address{Dept. of Physics
University of California at Los Angeles
Los Angeles, CA 90095}
\date{\today}
\maketitle
\tightenlines
\widetext
\advance\leftskip by 57pt
\advance\rightskip by 57pt

\begin{abstract}

When holes are doped into an antiferromagnetic insulator they form a slowly 
fluctuating array of ``topological defects'' (metallic stripes) in which the 
motion of the holes exhibits a self-organized quasi one-dimensional electronic 
character. The accompanying lateral confinement of the intervening
Mott-insulating regions induces a spin gap or pseudogap in the environment of
the stripes. We present a theory of underdoped high temperature 
superconductors and show that there is a {\it local} separation of spin and 
charge, and that the mobile holes on an individual stripe acquire a 
spin gap via pair hopping between the stripe and its environment; {\it i.e.} 
via a magnetic analog of the usual superconducting proximity effect. In this 
way a high pairing scale without a large mass renormalization is established 
despite the strong Coulomb repulsion between the holes. Thus the 
{\it mechanism} of pairing is the generation of a spin gap in 
spatially-confined {\it Mott-insulating} regions of the material in the 
proximity of the metallic stripes. 
At non-vanishing stripe densities, Josephson coupling between stripes produces 
a dimensional crossover to a state with long-range superconducting phase 
coherence.  This picture is established by obtaining exact and well-controlled 
approximate solutions of a model of a one-dimensional electron gas in an 
active environment. An extended discussion of the experimental evidence 
supporting the relevance of these results to the cuprate superconductors is
given.


\end{abstract}

]

\narrowtext
\tightenlines

\section{Introduction}

Superconductivity in metals is the result of two distinct quantum
phenomena, pairing and long-range phase coherence. In conventional 
homogeneous superconductors the phase stiffness is so great that
these two phenomena occur simultaneously. On the other hand, in granular 
superconductors and Josephson junction arrays, pairing occurs at the bulk 
transition temperature of the constituent metal, while long-range phase 
coherence occurs, if at all, at a much lower temperature characteristic of 
the Josephson coupling between superconducting grains.   High temperature
superconductivity\cite{BM} is hard to achieve, even in theory, because it 
requires that both scales be elevated simultaneously--yet they are 
usually incompatible.
Consider, for example, the strong-coupling limit of the negative $U$ Hubbard 
model \cite{negU} or the Holstein model \cite{holst}.  
Pairs have a large binding energy but, typically, they Bose condense at a 
very low temperature because of the large effective mass  of a tightly 
bound pair. (The effective mass is proportional to $|U|$ 
in the Hubbard model and is exponentially large in the Holstein model.)
A similar issue arises if the strong pairing occurs at specific locations
in the lattice (negative-$U$ centers); in certain limits this problem may be
mapped into a Kondo lattice\cite{coleman}, which displays heavy-fermion 
behavior.

A second problem for achieving {\hty} is that strong effective attractions,
which might be expected to produce a high pairing scale, typically lead to 
lattice instabilities, charge or spin density wave order, or two-phase
(gas-liquid or phase separated) states\cite{inhomogeneous}.
Here the problem is that the system either becomes an insulator or, if it 
remains metallic, the residual attraction is typically weak.
In the neighborhood of such an ordered state there is a low-lying collective 
mode whose exchange is favorable for superconductivity, but the superconducting 
transition temperature is depressed by vertex corrections \cite{jrs} and also
because the density of states may be reduced by the development of a 
pseudogap.

A third (widely ignored) problem is how to achieve a high pairing scale at 
all in the presence of the repulsive Coulomb interaction, especially in
a doped Mott insulator in which there is poor screening. A small
coherence length (or pair size) implies that neither retardation, nor a
long-range attractive interaction is effective in overcoming the bare 
Coulomb repulsion. Indeed, in the {\hts}, angle resolved 
photoemission spectroscopy\cite{arpes} (ARPES) suggests that the energy gap
(and hence the pairing force) is a maximum for holes separated by one lattice 
spacing, where the bare Coulomb interaction is very large.

\begin{figure}
\epsfysize=3.5in
\epsfbox{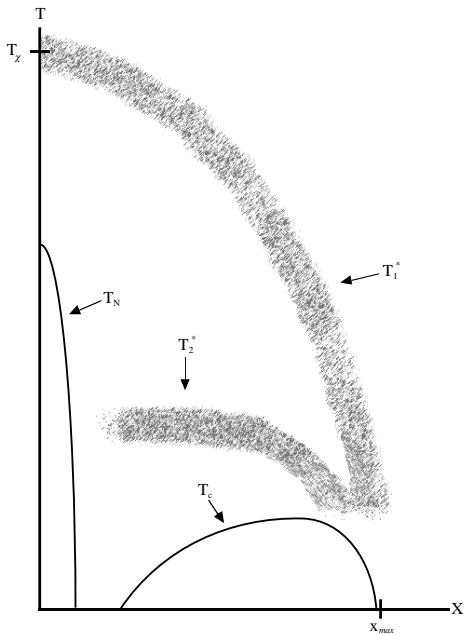}
\caption{
Theoretical sketch of the
phase diagram for a high temperature superconductor in the
doping-temperature plane.  The solid lines represent phase transitions
and the shaded areas crossovers.  $T_N$ marks the transition to
an antiferromagnetically ordered insulating state, and $T_c$ the
transition to the superconducting state.  $T_1^*$ marks the crossover
temperature at which charge inhomogeneities (stripes) form and
correspondingly local antiferromagnetic correlations develop in
the insulating regions;  the present paper is primarily concerned with
the region between $T_1^*$ and somewhat above $T_c$, where the developing 
correlations are primarily 
confined to the neighborhood of an individual stripe.  $T_2^*$ marks
the temperature scale at which a spin gap develops in the 1DEG,
and correspondingly the local superconducting susceptibility begins
to diverge.  Here, $T_{\chi}$, which is approximately $1/2$ the
antiferromagnetic exchange energy, marks the temperature at which
the antiferromagnetic correlation length in the undoped antiferromagnet
is equal to two or three lattice constants.  For further discussion, especially
concerning the experimental justification for this figure, see
Sec. IXC.}
\label{fig1}
\end{figure}

In short, superconductivity typically occurs at low temperatures: if 
any attractive interaction is weak the pairing energy is small; if it is
strong the coherence scale is suppressed or the system is otherwise unstable. 
When this is coupled with the problem presented by the Coulomb force in
a doped Mott insulator, the occurrence of high temperature
superconductivity in the cuprate perovskites is even more remarkable.
Indeed, there is evidence\cite{nature,badmetal,doniach} 
that these materials live in a region
of delicate balance between pairing and phase coherence:  
in ``underdoped'' and ``optimally doped''materials, the onset of 
superconductivity is controlled by phase coherence, and occurs well below the 
pairing temperature, while in ``overdoped'' materials 
pairing and phase coherence take place at more or 
less the same temperature, as in more conventional superconductors.
(See Fig. \ref{fig1}.) If we accept this
point of view, then we can approach the problem of understanding the
mechanism of high temperature superconductivity from the underdoped side
by addressing three separate questions:  i)  What gives rise to the large
temperature scale for pairing, or in other words for superconductivity 
on a local scale?  ii)  How can the system avoid the detrimental effects of 
strong pairing on global phase coherence? ({\it i.e.} large mass 
renormalizations.)  iii) How can {\hty} with a short coherence length coexist
with poor screening of the Coulomb interaction? 

Here we shall argue that the {\hts} resolve these 
problems in a unique manner: 1)  The tendency of an antiferromagnet
to expel holes\cite{spbag} leads to the formation of hole-rich and hole-free  
regions\cite{ekl}. For neutral holes this leads to a uniform instability 
(phase separation)\cite{ekl} but, for charged holes, the competition with
the long-range part of the Coulomb interaction generates a 
dynamical {\it local} charge inhomogeneity, in which the mobile holes are 
typically confined in ``charged stripes'', separated by elongated regions 
of insulating antiferromagnet\cite{comp,ute,spherical}.  This self-organized
collective structure, which we have named {\it topological doping}\cite{topo},
is a general feature of doped Mott insulators, and it produces a locally 
quasi one-dimensional electronic character since, the electronic coupling 
between stripes falls exponentially with the distance between them
\cite{zimanyi}.
2)  In a locally-striped structure, there is separation of spin and charge,
as in the one-dimensional electron gas\cite{LE} (1DEG). Hence ``pairing'' is 
the formation of a spin gap, while
the superfluid phase stiffness ({\it i.e.} the superfluid density
divided by the effective mass) is a property of the collective charge
modes\cite{strong,ZKL,KZ}.   
3) A large spin gap (or spin pseudogap)
arises naturally in a spatially-confined, hole-free 
region, such as the medium between stripes. This effect is 
well documented  for spin ladders\cite{ladder}, and for spin chains
with sufficient frustration\cite{CuGe,mg}. The important point is
that the spin gap does not conflict with the Coulomb interaction
since the energetic cost of having localized holes in Cu $3d$ 
orbitals has been paid in the formation of the material. 
4) The spin degrees of freedom of the 1DEG acquire a spin gap by pair 
hopping between the stripe and the antiferromagnetic environment.
(Single particle tunnelling is irrelevant\cite{tunnel}.)
At the same time, because of the local separation of spin and charge,
the spin-gap fixed point is stable even in the presence of strong Coulomb
interactions, and there is no mass renormalization to depress the onset of 
phase coherence, so the superconducting susceptibility diverges strongly
below this temperature\cite{diverge}.

In summary, the ``mechanism'' of high temperature superconductivity is a
form of magnetic  proximity effect in which a spin gap is generated in
{\it Mott-insulating} antiferromagnetic regions through spatial confinement by 
charge stripes, and communicated to the stripes by pair hopping. The mobile 
holes on the stripes have the large phase stiffness required for a high 
superconducting transition temperature.

The formation of a spin gap in the 1DEG may be regarded as a pairing of 
``spinons'', {\it i.e.} the neutral, spin-1/2 soliton excitations which 
occur in the low energy spectrum of the 1DEG and a number of one-dimensional
quantum antiferromagnets. Indeed, local inhomogeneity provides a realization 
of some of the earlier ideas\cite{spinliq} 
involving spin-charge separation in the {\hts} and the concept of a spin 
liquid, by which we mean a quantum disordered system 
({\it i.e.} with unbroken spin-rotation symmetry)
which supports spinons in its physical spectrum.
However, we emphasise that previous ideas relied on a
putative {\it two-dimensional} spin-liquid fixed point, while here we are 
dealing with a {\it locally} one-dimensional system, for which 
it is well established\cite{LE,ZKL} that separation 
of spin and charge \cite{LE}  occurs generically, and there exists a ``paired
spin-liquid'' phase, {\it i.e.} a spin-liquid with
a finite gap or pseudogap in the spinon spectrum. 
(See discussion in Appendix C.)
In the strictest sense then, both are intermediate-distance effects\cite{laughnew} 
which occur below a dimensional-crossover scale to 
two (or three) dimensional physics.  

We thus view the emergence
of high temperature superconductivity as a three-stage process, which
can be described in renormalization group language in terms of the 
influence of three fixed points.  At high temperatures, the ``avoided
critical phenomena''\cite{spherical} associated with frustrated phase
separation, govern the emergence of the self-organized, quasi
one-dimensional structures.  At intermediate temperatures, 
the one-dimensional {\it paired spin liquid} fixed point
controls the pairing scale, and the growth of local superconducting
(and CDW) correlations.  Finally, at low temperatures, a two (or three) 
dimensional fixed point determines the long-distance physics and
the ultimate superconducting or insulating behavior of the system.

Our proposed mechanism implies the existence of two
crossover scales above $T_c$ in underdoped materials, as shown in
Fig. 1:  a high temperature scale, at which local stripe order and
antiferromagnetic correlations develop, and a lower temperature at
which local pairing (spin gap)
and significant superconducting correlations appear on individual charge
stripes.  $T_c$ itself, is then determined by the Josephson coupling
between stripes, {\it i.e.} by the onset of global phase coherence
\cite{nature}.

The local charge inhomogeneity which is a central feature of our model has
substantial support from experiment.
In the past few years charge ordering has been discovered in a number of 
layered oxides,
such as {\LSNiO}\cite{jtni} and La$_{0.5}$Sr$_{1.5}$MnO$_4$\cite{ben},
and there is considerable experimental evidence showing 
that the {\hts} display a coexistence of superconductivity and charge 
inhomogeneity. In
particular, the efficient destruction of the antiferromagnetic order
\cite{nrev} of the 
parent insulating state is a consequence of {\it topological doping} 
\cite{topo}, in which
the mobile holes form metallic stripes that are antiphase domain walls for
the spins. The stripes may be ordered\cite{tranq} (as in {\LNSCO}), dynamically 
fluctuating \cite{tranq,Tvseps} (as in optimally-doped {\LSCO}), or pinned and 
meandering \cite{borsa} (as in lightly doped {\LSCO}). Thus, we consider the 
existence of local metallic stripes (at least in the {\LCO} family of high 
temperature superconductors) to be an experimental fact. Evidence of specific
charge fluctuations in any family of cuprate superconductors suggests that 
they are an important ingredient in the theory of {\hty}. However neutron 
scattering data\cite{sternlieb} also suggest that there are similar, but more 
disordered, structures\cite{spherical} in underdoped {\YBCO}. An
analysis of ARPES experiments on {\BSCCO} leads to a similar 
conclusion\cite{markku}.
		     
The systematics of phase fluctuations \cite{nature}, mentioned 
above, strongly suggests that  pairing on a high energy scale does not
require interaction between metallic charge stripes, although T$_c$ is certainly 
controlled by the Josephson coupling required to establish phase coherence 
for an array of stripes. Consequently, it should be possible to understand 
the mechanism of pairing from the behavior of a single stripe, modelled as a 
1DEG coupled to the various low-lying states of an insulating environment. 
A complete discussion of this problem is a substantial generalization of the 
theory of the one-dimensional electron gas \cite{Boso1D} which will be 
considered more completely in 
a subsequent publication \cite{ekz}. Here it will be shown that, for the 
{\hts}, the most important process is the hopping of a pair of holes from
the stripe into the antiferromagnetic environment, which also may be regarded
as a coherent form of transverse stripe fluctuation. It will be shown that
the stripe develops a spin gap which, in this model, corresponds to pairing 
without phase coherence. We consider two situations: a) the
antiferromagnetic environment has a pre-existing spin gap or spin pseudogap 
because of its finite spatial dimensions\cite{ladder} and b) pair hopping 
produces a spin gap in both the stripe and the environment. In the first case, 
we find that an induced spin gap in the 1DEG and the consequent 
divergent superconducting fluctuations are a robust consequence of the
coupling to the environment. The second case requires a sufficiently 
strong (and possibly unphysical) Coulomb interaction between holes on the 
stripe and holes in the environment for pair tunnelling to be relevant.

Although the existence of two distinct regions, the stripe and the
antiferromagnetic environment, provides a potential
escape from some of the limitations
on the superconducting transition temperature T$_c$,
it is not {\it a priori} obvious that a large mass renormalization 
can be avoided.
Indeed, the model we shall study is closely related to Kondo lattice
models\cite{coleman}, for which heavy-fermion behavior or large mass 
renormalization is the {\it primary} consequence of the strong interactions.  
However we find that, for stripes in an antiferromagnet (as for 
{\it one-dimensional} Kondo and orbital Kondo lattice models 
\cite{orbitalK,oron}), the analog of heavy-fermion physics is reflected 
solely in the the spin degrees of freedom while for the charge modes, and 
hence the superfluid phase stiffness, the mass is not renormalized!

In some respects, what we are doing is analogous to working out the
renormalization of the electron self energy by the coupling to phonons.
However, the calculation is more complicated because, here, the elementary 
objects are strings of charge (stripes) in a polarizable medium that
profoundly influences their internal structure. Fluctuating stripes are 
of finite length but the solution of the infinite 1DEG may be used if they 
are longer than the spin gap length scale, which is a few lattice 
spacings\cite{ladder}.

Of course, at higher hole concentrations, the  calculation must be modified 
to take account of the interaction between the stripes, especially to obtain
long-range superconducting order. In general terms, it is fairly straightforward 
to see how global superconductivity arises in a system with a small but
finite density of ordered or slowly-fluctuating stripes, as found 
in underdoped members of the {\LSCO} family of superconductors.
Indeed, an analysis of neutron scattering and thermodynamic data 
for underdoped and optimally doped {\LSCO}\cite{Tvseps} suggests that
T$_c$ is proportional to the product of the Drude weight of the holes 
on a stripe and the stripe concentration c$_s$.  

An interesting feature of our model is the interplay between the 
short-distance physics associated with the fluctuating stripes and the
ultimate long-range order that is established in a given material. We 
shall show that both superconducting and charge density wave correlations 
develop on a given stripe. However, they compete at longer length scales,
although they may coexist in certain regions of the phase diagram. Also it
follows from general principles that, locally, the singlet superconducting 
order parameter will be a strong admixture of extended-$s$ and 
d$_{x^2-y^2}$ states.  Ultimately, in tetragonal materials, the order 
parameter must have a pure symmetry, but the way in which it emerges from the
short-distance physics is very different from more conventional routes.

This paper is quite long and, in parts, rather technical.  It addresses the
purely theoretical problem of constructing and solving a general model of a 
1DEG in an active environment.  At the same time, we wish to report progress 
on the key problem of understanding the mechansim of {\hty} in the cuprate 
superconductors. To compensate, we have attempted to make the various sections as
self-contained as possible, and to indicate which sections can be skipped by 
the reader with a more focussed interest in the problem. 

A rather general model of the interactiong 1DEG in an active environment is 
introduced in Sec. II.  The model is bosonized in Sec. III, and
various formal transformations that are useful for later analysis are
described;  this section also contains a discussion of
which of the allowed interactions in the model are unimportant for
our purposes, and so can be ignored.  In Sec. IV,  we define
a simplified ``pseudospin'' model of the charge excitations of the environment, 
and argue that it exhibits the same low-energy physics as the general model.
Sec. V contains a discussion of exact results for the zero temperature
properties of the pseudospin model, which among other things
exhibits the spin-gap proximity effect, and the generation of a paired spin 
liquid state of the 1DEG, even in the presence of
arbitrarily-strong forward scattering.
Section VI reports the results
of a controlled approximate solution of the pseudospin model
for a wide range of temperatures and coupling constants;  in particular,
various crossover temperatures to spin-gap behavior
are identified, and their dependence
on the interactions in the model are determined.  In Sec. VII, we
return to the problem of the charge degrees of freedom of the 1DEG, and
consider the effects of umklapp scattering in conditions of near commensurability,
and the effects of an externally applied potential.  
In Sec. VIII, we digress slightly to consider the effects of a
``spin-gap center'' on the local properties of a Fermi liquid.
Finally, in Sec. IX, we summarize our results and discuss experimental
implications and predictions for the high temperature superconductors. In 
this section, we also suggest some numerical 
calculations to test the major ideas. The reader who is primarily interested 
in a discussion of results may skip directly to Sec. IX. 
In addition,  Appendix A recasts
some of the present discussion in the familiar language of the
perturbative renormalization group for the 1DEG, Appendix B contains
an analysis of the symmetries of the model, and an explicit construction
of the non-local order parameter which characterizes ``local pairing'',
and Appendix C discusses the precise nature of the paired-spin-liquid state,
and gives concrete examples of model systems which exhibit this state.

\section{The 1DEG in an Active Environment}

\subsection{The problem and the solution strategy}

It has long been realized that the low energy properties of a one
dimensional electron gas (1DEG),
and indeed of a wide variety of other interacting
one dimensional systems, are equivalent to those of the simplest field theory
of interacting electrons, characterized by a small
number of potentially relevant interactions between electrons at the
Fermi surface.  In this section we address the problem of a
1DEG in an ``active'' environment, one that possesses its
own low-energy excitations which couple to the 1DEG, but is insulating 
so that the electrons of the 1DEG may make excursions into the environment, 
but ultimately return. The environment in which we are interested is 
antiferromagnetic, so it may have low-energy spin excitations.  It will
also have low-energy charge excitations in which holes make excursions from 
the metallic stripe into the environment. Their energy is low because 
frustrated phase separation, which generates metallic stripes in the first 
place, involves a delicate balance of Coulomb and magnetic energies.

This problem can be addressed in several distinct ways.  In the present paper,
we make extensive use of a renormalization group strategy involving exact 
solutions of solvable models, 
together with a sophisticated approximate calculation, in which the
fluctuations of the 1DEG and the environment are solved exactly, but the 
coupling between them is treated in a mean-field approximation.  We also 
give physical estimates of the values of the various coupling constants that 
enter the model, and present strong physical arguments to show that the 
physical systems of interest will lie in the ``basin
of attraction'' of the strong-coupling fixed point that governs the
behavior of the solvable models.
In Section IX, we will also outline some simple one-dimensional lattice
models which are amenable to numerical solution, and are expected to
exhibit the mechanism described in this paper.

\subsection{The general model}

To begin with, we consider a very general model of a 1DEG coupled to an
environment. The initial form of the model is microscopically realistic.
It will be assumed that the environment itself is a one dimensional system
with a charge gap  (since it is an insulating matrix) which may or may not
have a spin gap.
We thus consider the Hamiltonian to be of the form
\be
H=\int_{-\infty}^{\infty} dx
\big[ {\cal H}_{1DEG} + {\cal H}_{env} + {\cal H}_{int} + {\cal H}_{coul}
\big].
\label{eq:envham}
\ee

The bare Hamiltonian density
of the 1DEG is
\be
{\cal H}_{1DEG}={\cal H}_0 + {\cal H}_1.
\label{eq:H1DEG}
\ee
Here ${\cal H}_0$ is the Hamiltonian of a non-interacting 1DEG,
which in the continuum limit can be written (with $\hbar=1$) as
\ba
{\cal H}_0= & & i v_F\sum_{\sigma} \left[\psi^{\dagger}_{1,\sigma}
\partial_x \psi_{1,\sigma} - \psi^{\dagger}_{2,\sigma}
\partial_x \psi_{2,\sigma}\right] \nonumber \\
& & -\mu
\sum_{\alpha,\sigma}\left
[\psi^{\dagger}_{\alpha,\sigma}(x) \psi_{\alpha,\sigma}(x) \right]
\label{eq:H0}
\ea
where $\psi^{\dagger}_{\alpha,\sigma}(x)$ creates an electron with z
component
of spin $\sigma$ on the right or left moving branch of the Fermi
surface for $\alpha=1$ or $2$ respectively.  Here, we have made a
Galilean transformations to shift the Fermi points to $k=0$; factors
involving the Fermi wave vector $k_F$ will be shown explicitly. ${\cal H}_1$
incorporates the electron-electron interactions within the 1DEG and has the
continuum form \cite{Boso1D}
\ba
{\cal H}_1 =& &  g_2\sum_{\sigma,\sigma^{\prime}}
 \psi^{\dagger}_{1,\sigma}
\psi^{\dagger}_{2,\sigma^{\prime}}\psi_{2,\sigma^{\prime}}
\psi_{1,\sigma}
 \nonumber \\
& &  +g_1\sum_{\sigma,\sigma^{\prime}}
\psi^{\dagger}_{1,\sigma}
\psi^{\dagger}_{2,\sigma^{\prime}}\psi_{1,\sigma^{\prime}}
\psi_{2,\sigma} \nonumber \\
& & + g_3
\big [\psi_{1,\uparrow }^{\dagger}\psi_{1,\downarrow }^{\dagger}
\psi_{2,\downarrow }\psi_{2,\uparrow }e^{i(4k_F-G)x} +  H.c.\big].
\label{eq:H1}
\ea
Here $G$ is a reciprocal lattice vector and $g_3$ is the coupling constant
for umklapp scattering.  When the 1DEG is incommensurate $(4k_F \neq G)$,
the rapid phase oscillations in the term proportional to $g_3$ render it
irrelevant in the renormalization group sense. However, near to
commensurability, this term is responsible for the fact that the Drude weight
is proportional to the density of doped holes, as we shall see.
Typically, it will be assumed that the
interactions are repulsive ($g_1, g_2, g_3 > 0$) although they may undergo
significant renormalization by the coupling of the 1DEG to the {\it high
energy} excitations of the antiferromagnetic environment (which we do not
consider explicitly).
The parameters that describe
the 1DEG are thus the Fermi velocity, $v_F$, the chemical
potential, $\mu$, the three coupling constants $g_i$, and the
``incommensurability'', $4k_F-G$.
It should be emphasised that this is a very general representation of the
low-energy physics of a stripe in a CuO$_2$ plane, and all details of the
original microscopic model are contained in the values of the coupling
constants $g_i$.

We have in mind the low-density limit of a stripe phase in which the
Coulomb interaction on a given stripe is screened by the motion of charge
on neighboring stripes, and so does not make a singular contribution to
the forward scattering interaction, $g_2$. Thus, for
the time being, we will neglect the term ${\cal H}_{coul}$, although it will
ultimately play a role in the dynamics of the superconducting phase
\cite{badmetal}.

Because the physics of interacting systems in one dimension is
ultimately so constrained, it is possible to model the Hamiltonian density of
the environment as a second (distinct) interacting one dimensional electron
gas. The Hamiltonian ${\cal H}_{env}$ has the same form as in Eqs. (\ref{eq:H0})
and (\ref{eq:H1}), except that fields and parameters will be marked with a
super-tilde. However there are several
important differences in the parameters of the Hamiltonian:
1) The environment is a Mott insulator. Consequently there is a strong
commensurability energy ($4{\tilde k}_F=\tilde G$ and ${\tilde g}_3$ is large),
which produces a gap in the the charge degrees of freedom of the environment.
This also implies that ${\tilde k}_F$ is different from $k_F$.
2) Because of the frustration of the motion of holes in an antiferromagnet,
\cite{trugman}
the propagation velocity $\tilde v_c$ for charge excitations in the
environment is much smaller than the corresponding velocity in the 1DEG.
This is the primary manner in which the driving force for phase separation
\cite{ekl} and stripe formation \cite{ute,spherical} appears in the model.
3) We shall consider three possibilities for the spin degrees of freedom of the
environment, one in which there are gapless magnon-like excitations,
and two in which there is a spin gap:  a)  The gapless state is realized
by considering the model with $\tilde g_1 >0$, in which case the
environmental spin excitations are those of an antiferromagnetic
spin-1/2 Heisenberg chain.  b)  A spin
gap can occur with an accompanying spontaneous
breaking of translational (chiral)
symmetry (See Appendix B),
which is realized by simply taking $\tilde g_1 <0$, in which case the
environmental spin excitations are those of a spin-1/2 Heisenberg chain
with competing nearest and next-nearest neighbor antiferromagnetic
interactions, {\it e.g.} the Majumdar-Ghosh model.\cite{mg}  c)
A spin-gap can occur without any accompanying broken symmetry, in the
manner of the antiferromagnetic two leg, spin 1/2 Heisenberg 
ladder\cite{ladder};
to model this system, we need to add a backscattering term to the
environmental Hamiltonian 
(of the same form as $H_e$ in Eq. (\ref{eq:He}), below),
although a better description
can be attained in the bosonized form of the Hamiltonian, as discussed
below.  For our purposes, there is no significant difference in the
implications of the two types of environmental spin gap, so
for simplicity, we will perform our calculations for the case
in which the spin gap is induced by a negative $\tilde g_1$, and
will use language to describe the physics that (properly) does not
distinguish the two types of environmental spin gap.

Using well known
results for the 1DEG, it is possible to express these coupling
constants in terms of the physical variables which define the
excitation spectrum of the environment: the spin and charge velocities,
${\tilde v}_s$ and ${\tilde v}_c$, the charge gap ${\tilde \Delta}_c$ and
the spin gap (if one exists) ${\tilde \Delta}_s$, and the charge and spin
correlation exponents (defined below), $\tilde K_c$ and
$\tilde K_s$. Since the environment is an insulator, we will always
assume that ${\tilde \Delta}_c$ is large.
We also must include the energy $\epsilon$ to
transfer charge from the 1DEG to the environment.  For the case of
``p-type'' doping, in which $\tilde\mu$ lies in the lower half of the
environmental gap, $\varepsilon/2
\equiv{\tilde \Delta}_c - [\tilde\mu-\mu]$ is the bare energy required to
remove a quantum of charge
from the environment and add it to the 1DEG.  We will be interested
in the case $0\le\varepsilon\ll  {\tilde \Delta}_c$.

Finally, we consider the coupling between the 1DEG and the environment, for
which spin-rotational invariance and conservation of momentum along the
stripe direction severely limit the number of possible relevant interactions.
Since the Fermi wave vector of the 1DEG is incommensurate with the
wave vector of any low energy excitation of the environment,
we can neglect, as irrelevant, terms which transfer momentum $\pm k_F$ or
$\pm 2k_F$ between the 1DEG and the environment.  For example
there are no low energy single-particle hopping processes, even though,
at the microscopic level, one might expect them to have the
largest coupling term. Such processes are included implicitly as virtual
intermediate states in constructing the effective low energy Hamiltonian.
(We will return to this point briefly in the following section.)
With this in mind, the most general form of the interaction Hamiltonian
density, {\it i.e.} which keeps all potentially relevant terms, is
\ba
{\cal H}_{int}  = && J_s {\vec j}_s \cdot {\vec{\tilde j}}_s +
V_s {\vec S}\cdot {\vec{\tilde S}} \nonumber\\
&& +  J_c  j_c{\tilde j}_c + V_c \rho{\tilde \rho}  \nonumber\\
&& +   {\cal H}_{pair},
\ea
where the small momentum transfer couplings involve
the long-wavelength density
fluctuations relative to the background charge density $\rho_0$
\be
\rho(x) - \rho_0 = \sum_{\sigma}
\big[\psi_{1,\sigma}^{\dagger}\psi_{1,\sigma} +
\psi_{2,\sigma}^{\dagger}\psi_{2,\sigma} \big],
\ee
the {\it bare} charge-current operator
\be
j_c(x)=\sum_{\sigma}
\big[\psi_{1,\sigma}^{\dagger}\psi_{1,\sigma} -
\psi_{2,\sigma}^{\dagger}\psi_{2,\sigma} \big],
\ee
the long-wavelength spin density operator
\be
{\vec S}(x)=\sum_{\sigma,\sigma'}
\big[ \psi_{1,\sigma}^{\dagger}
{\vec \sigma}_{\sigma,\sigma'}\psi_{1,\sigma'} +
\psi_{2,\sigma}^{\dagger}
{\vec \sigma}_{\sigma,\sigma'}\psi_{2,\sigma'} \big],
\ee
and the {\it bare} spin-current operator
\be
{\vec j}_s(x)=\sum_{\sigma,\sigma'}
\big[ \psi_{1,\sigma}^{\dagger}
{\vec \sigma}_{\sigma,\sigma'}\psi_{1,\sigma'} -
\psi_{2,\sigma}^{\dagger}
{\vec \sigma}_{\sigma,\sigma'}\psi_{2,\sigma'} \big].
\ee
The corresponding operators for the environment are defined by the same
equations, except that all quantities have a super-tilde.
Note that we have chosen to express ${\cal H}_{int}$ in terms of the
charge and spin current operators for the noninteracting system.
The other contribution to ${\cal H}_{int}$ is the pair transfer terms
\ba
{\cal H}_{pair}= &&  t_{sp} \big[P^{\dagger}{\tilde P} + {\rm H.c.}\big]
\nonumber \\
+ && t_{tp} \sum_{m=-1}^1\big[{ P}_m^{\dagger}{{\tilde P}}_m + {\rm H.c.}\big]
\label{eq:Hpair}
\ea
where for the 1DEG, $P^{\dagger}$ is the usual singlet pair creation operator,
\be
P^{\dagger}(x) \equiv
\frac 1 {\sqrt{2}}\left[\psi^{\dagger}_{1,\uparrow}(x)
\psi^{\dagger}_{2,\downarrow}(x) +
\psi^{\dagger}_{2,\uparrow}(x)
\psi^{\dagger}_{1,\downarrow}(x)\right],
\label{eq:pairop}
\ee
and $P_m$ are the componenets of the triplet pair creation operator,
\ba
 P^{\dagger}_1(x) & \equiv &  \psi_{1,\uparrow}^{\dagger}
\psi_{2,\uparrow}^{\dagger} \nonumber \\
P_0^{\dagger}(x) & \equiv &
\frac 1 {\sqrt{2}} \left[\psi^{\dagger}_{1,\uparrow}(x)
\psi^{\dagger}_{2,\downarrow}(x) -
\psi^{\dagger}_{2,\uparrow}(x)
\psi^{\dagger}_{1,\downarrow}(x)\right] \nonumber \\
P^{\dagger}_{-1} & \equiv &  \psi_{1,\downarrow}^{\dagger}
\psi_{2,\downarrow}^{\dagger}.
\label{eq:trippair}
\ea

\section{Bosonization of the Model}

In dealing with the problem of the 1DEG in an active environment,
it is useful to rewrite the model using the
standard boson representation of Fermi fields in one
dimension \cite{Boso1D}:
\begin{equation}
\psi^{\dagger}_{\lambda,\sigma}(x)= \frac {1} {\sqrt {2\pi a}}
exp\{i\Phi_{\lambda,\sigma}(x) \}
\label{eq:bosonrep}
\end{equation}
where
$\Phi_{\lambda,\sigma} =
\sqrt{\pi} [ \theta_{\sigma} (x) \pm \phi_{\sigma}(x) ]$
with ``-'' and ``+'' corresponding to $\lambda=1$ and $2$
respectively, $\theta_{\sigma}(x)= \int_{- \infty}^x dx' \Pi_{\sigma}(x')$,
and
$\phi_{\sigma}(x)$ and $\Pi_{\sigma}(x)$ are canonically conjugate Bose
fields, so that $[\phi_{\sigma}(x),\Pi_{\sigma'}(x')] = i \delta(x-x')$.
($\theta$ and $\phi$ are thus dual to each other in the usual
statistical mechanical sense of order and disorder variables.)
To take advantage of the separation of spin and charge \cite{Boso1D},
the Hamiltonian will be expressed in terms of a spin field,
$\phi_s(x) = [\phi_{\uparrow}- \phi_{\downarrow}]/ \sqrt{2}$, and a charge
field, $\phi_c(x) = [\phi_{\uparrow}+\phi_{\downarrow}]/\sqrt{2}$ and
their conjugate momenta:
$\Pi_s(x) = [\Pi_{\uparrow}- \Pi_{\downarrow}]/ \sqrt{2}$ and
$\Pi_c(x) = [\Pi_{\uparrow}+\Pi_{\downarrow}]/\sqrt{2}$.
The charge and spin density and current operators may be
written:
\ba
\rho(x) & = & -\sqrt{2 \over \pi} \ \partial_x \phi_c
\\
j_c(x) & = &  \sqrt{2 \over \pi} \ \Pi_c
\nonumber \\
S^z(x) & = & -\sqrt{1 \over 2 \pi} \ \partial_x \phi_s
\nonumber \\
S^{\pm}(x) & = & {1 \over  \pi a} \exp(\pm i \sqrt{2 \pi} \theta_s)
\cos [\sqrt{2 \pi} \phi_s]
\nonumber \\
j_s^z(x) & = & \sqrt{1 \over 2 \pi} \ \Pi_s
\nonumber \\
j_s^{\pm}(x) & = & {- i \over  \pi a} \exp(\pm i \sqrt{2 \pi} \theta_s)
\sin [\sqrt{2 \pi} \phi_s].
\ea

In terms of these variables, the Hamiltonians of the stripe, the environment,
and the small-momentum transfer coupling between the two may be written as a
sum of a charge-only part  and a spin-only part. However, the pair hopping
terms $H_{pair}$ introduces a coupling between spin and charge.
Thus the total Hamiltonian may be written
\be
{\cal H} = {\cal H}_c + {\cal H}_s + {\cal H}_{pair}.
\ee
We now consider the various contributions in turn.

\subsection{Spin Degrees of Freedom}

The general form of the spin Hamiltonian is
\be
{\cal H}_{s}  \equiv  {\cal H}_s^0 + {\cal H}_s^1 + {\cal H}_s^2
\label{eq:hs}
\ee
Here
\ba
{\cal H}_s^0 & = & \frac {v_s} 2
[ K_s \Pi_{s}^2 + \frac 1 {K_s} (\partial_x\phi_s)^2  ]  \nonumber \\
& + & \frac {\tilde v_s} 2  [ \tilde K_s \tilde \Pi_{s}^2 +
\frac 1 {\tilde  K_s} (\partial_x \tilde \phi_s)^2  ] ,
\label{eq:spin0}
\ea
\be
{\cal H}_s^1  =  \frac {2J_s} {\pi} \Pi_s(x)\tilde \Pi_s(x)
+ \frac {2V_s} {\pi}
\partial_x \phi_s(x) \tilde \partial_x \phi_s(x),
\label{eq:spin1}
\ee
and
\ba
{\cal H}_s^2   =  & & \frac {2 g_1} {(2 \pi a)^2}
\cos \big [\sqrt{8 \pi} \phi_s \big ]
\label{eq:spin2}
\\
+ &. &  \frac {2 \tilde g_1} {(2 \pi a)^2}
\cos \big [\sqrt{8 \pi} \tilde \phi_s \big ]
\nonumber \\
 + &  & \frac {V_s} {2 \pi a} \cos \big [\sqrt{2 \pi}(\theta_s-
\tilde \theta_s)\big ]
\cos \big [\sqrt{2 \pi} \phi_s \big]  \cos\big[\sqrt{2 \pi} \tilde \phi_s
\big ]
\nonumber \\
 + & & \frac {J_s} {2 \pi a} \cos \big [\sqrt{2 \pi}(\theta_s- \tilde \theta_s)
\big ]
\sin \big [\sqrt{2 \pi} \phi_s\big]  \sin\big[\sqrt{2 \pi} \tilde \phi_s \big ]
\nonumber
\ea
Here $v_s$ is the spin-wave velocity and $K_s$ is the critical 
exponent\cite{expdef} that specifies the location on a line of fixed points. 
Also $v_s$ is given by $v_s = 2v_F K_s/ (K_s^2 + 1)$.
In the absence of coupling between the stripe and the environment,
the Hamiltonian is known to be correct for weak or strong coupling and for
different forms of short-distance or high-energy cutoff \cite{Boso1D},
although it may be necessary to perform some form of global renormalization
to determine $K_s$ from the parameters of the initial Hamiltonian.
For weak coupling, $K_s$ is related to the bare Fermi velocity
$v_F$ and coupling constants as
$K_s = \sqrt{ \frac {2 \pi v_F + g_1} {2 \pi v_F - g_1} }$.
For repulsive interactions ({\it i.e.} $g_1>0$) one finds $K_s > 1$.

For the case in which $\tilde g_1$ is negative and relevant,
in the renormalization group sense, there is a two fold degenerate ground 
state, corresponding to the classical values $\tilde\phi_s=0$ and 
$\tilde\phi_s=\sqrt{\pi/2}$. (See Appendix B.) To represent the
case in which there is an environmental spin gap without symmetry breaking, 
we should add a term proportional to $\cos[\sqrt{2\pi}\tilde\phi_s]$,
which arises in a microscopic system with two spins per unit cell, such as
a two-leg ladder\cite{freeze}. This term (which may be generalized to allow any even 
number of spins per unit cell) is always relevant for repulsive interactions, 
so it always leads to a spin gap. As we shall see shortly, the important 
point is that a spin gap of whatever origin implies a quenching of the 
fluctuations of $\tilde \phi_s$.  For a caveat on commensurability effects,
see Sec. VII.

\subsection{Charge Degrees of Freedom}

The general form of the charge Hamiltonian is
\be
{\cal H}_c = {\cal H}_c^0 + {\cal H}_c^1 + {\cal H}_c^2 ,
\label{eq:Hc}
\ee
where
\ba
{\cal H}_c^0 & = & \frac {v_c} 2
[ K_c \Pi_{c}^2 + \frac 1 {K_c} (\partial_x\phi_c)^2  ]  \nonumber \\
& + & \frac {\tilde v_c} 2  [ \tilde K_c \tilde \Pi_{c}^2 +
\frac 1 {\tilde  K_c} (\partial_x \tilde \phi_c)^2  ] ,
\label{eq:charge0}
\ea
\ba
{\cal H}_c^1 & = & \frac {2J_c} {\pi} \Pi_c\tilde \Pi_c
+  \frac {2V_c} {\pi}
\partial_x \phi_c \partial_x \tilde \phi_c,
\label{eq:charge1}
\ea
and
\ba
{\cal H}_c^2  = &&  \frac {2 g_3} {(2 \pi a)^2}
\cos \big [\sqrt{8 \pi} \ \phi_c - (4k_F-G)x \big ] \nonumber \\
+ &&  \frac {2 \tilde g_3} {(2 \pi a)^2}
\cos \big [\sqrt{8 \pi} \tilde  \phi_c \big ] - \tilde\mu \sqrt{{2 \over \pi}} \
\partial_x \tilde \phi_c .
\label{eq:charge2}
\ea
Here $v_c$ is the charge velocity and $K_c$ is the Luttinger liquid exponent
\cite{expdef}, with $v_c = 2v_F K_c/ (K_c^2 + 1)$.
For weak coupling, $K_c$ is related to the bare Fermi velocity
$v_F$ and coupling constants as
$K_c = \sqrt{ \frac {2 \pi v_F + g_c} {2 \pi v_F - g_c} }$ ,
where $g_c = g_1 - 2 g_2$. For repulsive interactions
$0 < K_c < 1$ ({\it i.e.} $g_c < 0$).

\subsection{Spin-Charge Coupling}

Pair hopping between the stripe and the environment, as given by
${\cal H}_{pair}$, destroys the separation of spin and charge and is the
driving force for much of the interesting physics. Its bosonized form is
given by
\ba
{\cal H}_{pair} = && \bigg ({t_{sp} \over \pi^2a^2}\bigg )
\cos[\sqrt{2\pi}(\theta_c-{\tilde\theta_c})]
\cos[\sqrt{2\pi}\phi_s] \cos[\sqrt{2\pi}{\tilde\phi_s}] \nonumber \\
 + &&  \bigg ({t_{tp} \over \pi^2a^2}\bigg ) \bigg\{
\cos[\sqrt{2\pi}(\theta_c-{\tilde\theta_c})]
\cos[\sqrt{2\pi}(\theta_s-{\tilde\theta_s})] \nonumber \\
&& - \cos[\sqrt{2\pi}(\theta_c-{\tilde\theta_c})]
\sin[\sqrt{2\pi}\phi_s] \sin[\sqrt{2\pi}{\tilde\phi_s}] \bigg\}.
\label{eq:pairhop}
\ea

\subsection{Which Terms Are Unimportant?}

The general model has numerous coupling constants, and so, for much of this 
paper, we focus on the terms that are most important for our purposes, and 
set the others to zero.  Specifically, we drop those terms which are, in the
renormalization group sense, irrelevant at the paired spin liquid fixed point.
This argument simply shows that dropping these terms is
self consistent.  However, given
the nature of the antiferromagnetic environment, there are strong arguments 
to show that these terms also are physically irrelevant,
{\it i.e.} that the physical system lies in the basin of 
attraction of the paired spin liquid fixed point.

To begin with, we examine the magnetic interactions, $J_s$ and $V_s$ in
${\cal H}_{int}$:  these terms represent the interaction between the
{\it ferromagnetic} fluctuations in the two sub-systems.  Since we are primarily
interested in antiferromagnetic systems, we do not expect these terms ever
to be important.  Of course, in the paired spin liquid state, or more generally
in the presence of any sort of environmental spin gap, this can be seen directly
from their dependence on $\tilde\theta_s$, which means that the
corresponding correlation functions decay exponentially with distance
or time, and are thus trivially irrelevant.  The triplet pair-tunnelling
term similarly depends on $\tilde\theta_s$, and correspondingly triplet pairing
is generally expected to be important only in nearly ferromagnetic systems.
Therefore, on both clear physical, and formal renormalization group
grounds, it is safe to simplify our further discussion by taking
\be
J_s=V_s=t_{tp}=0,
\label{eq:ignore1}
\ee
unless explicitly stated otherwise.  

Thus, in the case where there is strong incommensurability between the
values of $k_F$ in the two subsystems, and neither has significant
ferromagnetic fluctuations, the only important interactions between 
the 1DEG and the environment are $t_{sp}$, $V_c$, and $J_c$.

Away from half filling, the renormalization of the umklapp scattering
coupling constant $g_3$ is cut off by the incommensurability\cite{ELP,ZKL},
and for some purposes it may be dropped.
However, this does not mean that umklapp scattering is unimportant for the
low energy physics. Doping of holes into the Mott insulating state in one
dimension creates soliton excitations\cite{ELP,soldop} in the charge density
with a mass governed by $g_3$. There is a ``doped-insulator'' region in
which these excitations control the Drude weight and the superfluid phase stiffness.
In our stripe model of the cuprates, {\hty} may occur within this region of
doping.

Finally, we address the non-linear term proportional to $g_1$ in
${\cal H}_s^2$ in Eq. (\ref{eq:spin2}).  For repulsive interactions,
{\it i.e.} for $K_s>1$, this term is perturbatively irrelevant, and the 
renormalization group flows go to the fixed point $g_1=0$ and $K_s=1$.
(See Appendix A.) Thus, so long as the bare interactions in the 1DEG are not 
too large, it is reasonable to use the fixed point values
\be
g_1=0 \ \ {\rm and} \ \ K_s=1
\label{eq:ignore2}
\ee
for the effective low energy theory.

\subsection{Unitary Transformation}

We now introduce a unitary transformation which will be used in a number of
ways to simplify the problem. The operator
\be
U_{\lambda}=\exp[-i \lambda \int dx \  \theta_c(x)\partial_x{\tilde\phi_c}(x)]
\label{eq:unitary}
\ee
has the effect of shifting the fields
\ba
U^{\dagger}_{\lambda}{\tilde \Pi_c}(x)U_{\lambda} &=&{\tilde \Pi_c}(x) +
\lambda \Pi_c(x)
\nonumber \\
U^{\dagger}_{\lambda}\tilde \phi_c(x) U_{\lambda} &=& \tilde \phi_c(x),
\nonumber \\
U^{\dagger}_{\lambda}{\Pi_c}(x)U_{\lambda} &=&{\Pi_c}(x)
\nonumber \\
U^{\dagger}_{\lambda}\phi_c(x) U_{\lambda} &=& \phi_c(x) - \lambda {\tilde
\phi_c}(x).
\label{eq:shift}
\ea
This transformation modifies the various charge interactions
\ba
V_c & \rightarrow & \Delta V_c = V_c-
{\pi \over 2} \frac {\lambda v_c} {K_c} \nonumber \\
J_c & \rightarrow & \Delta J_c = J_c-
{\pi \over 2} \lambda \tilde v_c \tilde K ,
\label{eq:deltaVc}
\ea
and the velocities and exponent parameters
\ba
v_c & \rightarrow & v_c\gamma \nonumber \\
K_c & \rightarrow & K_c\gamma \nonumber \\
\tilde v_c & \rightarrow & \tilde v_c\tilde \gamma \nonumber \\
\tilde K_c & \rightarrow & \tilde K_c/ \tilde\gamma \nonumber \\
\label{eq:deltaKc}
\ea
where
\ba
\gamma &=& \sqrt{1 + \frac {\lambda^2 \tilde v_c\tilde K_c} {v_cK_c} +
\frac {4 \lambda J_c} {\pi v_cK_c} } \nonumber \\
\tilde\gamma &=& \sqrt{1 + \frac {\lambda^2 v_c\tilde K_c} {\tilde v_cK_c} -
\frac {4\lambda V_c\tilde K_c} {\pi \tilde v_c} }.
\label{eq:gamma}
\ea

\subsubsection{Perturbative Relevance of Pair Hopping}

The transformation (\ref{eq:unitary}) diagonalizes the quadratic part of
the charge Hamiltonian ${\cal H}_c^0 + {\cal H}_c^1$  provided\cite{provided}
\ba
\lambda & = & \frac {2 V_c K_c} {\pi v_c} \nonumber \\
J_c & = & - \frac {{\tilde v}_c {\tilde K}_c K_c V_c} {v_c}.
\label{eq:lambda}
\ea

We are now in a position to discuss the perturbative relevancy of pair
hopping, which is the process that will generate a spin gap along the stripe.
Here we have in mind the initial stage of renormalization, in which degrees 
of freedom with energies large compared to the charge transfer energy, 
$\varepsilon$, are eliminated. Thus it is reasonable to determine
perturbative relevance relative to the quadratic piece of the Hamiltonian
\cite{relevance}. (See also Appendix A.) However other relevant pertubations, 
such as $\tilde g_3$, are important for the later stages of renormalization.
Substitution of  Eqs. (\ref{eq:lambda}) into Eqs. (\ref{eq:gamma}) gives
\ba
\gamma & = &
\bigg [ 1 - \frac {4 V_c^2} {\pi ^2} \frac {\tilde v_c \tilde K_c K_c}
{v_c^3}\bigg ]^{1/2}
\nonumber \\
\tilde \gamma & = &
\bigg [ 1 - \frac {4 V_c^2} {\pi ^2} \frac {\tilde K_c K_c}
{v_c \tilde v_c}\bigg ]^{1/2}.
\label{eq:gammadiag}
\ea
Then the singlet pair hopping operator ${\cal H}_{pair}$ is pertubatively 
relevant\cite{Boso1D} if the exponent
\be
\alpha_{sp} = {1 \over 4} \bigg ( \frac { \tilde \gamma} {\tilde K_c}
+ \frac{(1- \lambda)^2}{\gamma K_c} + \tilde K_s +  K_s \bigg ).
\label{eq:alpha}
\ee
satisfies $\alpha_{sp} < 1$, and perturbatively irrelevant otherwise.
Despite appearances, $\alpha_{sp}$ shares the property of the Hamiltonian
that it is symmetric under interchange of $\tilde K_c$ and $K_c$ when 
$\tilde v_c = v_c$.
If all interactions in the original model were set equal to zero,
then all of the $K's$ and $\gamma$'s would be equal to 1, so that  
$\alpha_{sp} = 1$, and pair hopping would be marginal.
Repulsive interactions within the stripe and the environment increase the
value of $\alpha_{sp}$, since they make $K_s, \tilde K_s \ge 1$ and
$K_c, \tilde K_c < 1$.
This is physically reasonable because repulsive interactions within the
stripe and the environment are unfavorable for pairing.

There are three effects which enhance the perturbative relevance of singlet 
pair hopping. First of all, it can be seen from Eqs. (\ref{eq:gammadiag}) and 
(\ref{eq:alpha}), that a repulsive $V_c$ decreases the value of $\alpha_{sp}$.
Physically, this occurs because the charge density in the environment
decreases in the vicinity of a pair in the 1DEG; thus it is easier for the
pair to hop. This effect is surely an important piece of the physics of
pair hopping and it provides a way in which the Coulomb repulsion is
favorable for pairing. But it cannot be the sole reason for the relevancy of
singlet pair hopping unless $V_c$ is greater than a suitable average of
$|g_c|$ and $|\tilde g_c|$.
As discussed in Appendix A, this can happen, in principle, if the character
of the screening is just right, but it seems to be an insufficiently robust
mechanism for a high temperature scale for pairing.

Secondly, the frustration of the motion of holes in an antiferromagnet implies
that the bare Fermi velocity $\tilde v_F$ of the environment is small, and 
hence $\tilde v_c$ is small, which depresses the value of $\tilde \gamma$
(Eq. (\ref{eq:gammadiag})) and the first contribution to $\alpha_{sp}$ in
Eq. (\ref{eq:alpha}).

Thirdly, if the environment has a preexisting spin gap, then
one should set $\tilde K_s=0$ in the expression for $\alpha_{sp}$;
this substitution makes singlet pair hopping
perturbatively relevant ({\it i.e.} $\alpha_{sp} < 1$)
for a wide range of the other parameters.
A slightly weaker form of this route occurs if the environment has a spin
pseudogap. For example it might have several gapped spin excitations and one 
gapless spin excitation, as in odd-leg ladders\cite{ladder}. 
Then the $\tilde K_s$-term in $\alpha_{sp}$ should have a
coefficient $w_s < 1$ equal to the weight  of the gapless excitation in the
pair hopping process. The elimination or reduction of $\tilde K_s$
in Eq. (\ref{eq:alpha}) is the perturbative
renormalization group manifestation of the proximity effect.

It is important to note that transverse fluctuations of the stripe,
together with the Coulomb interaction between holes on the
stripe and in the environment, increase the value of the superexchange
coupling along neighboring bonds perpendicular to the stripe\cite{ER}.
Clearly these processes decrease the value of $w_s$ and
are almost as effective as a full environmental spin gap for making pair
hopping perturbatively relevant.
Moreover the environment will vary along the length of a
fluctuating stripe, and singlet pair hopping may be relevant at some stripe
locations (``spin-gap centers'') and irrelevant at others, where it may be
neglected.  This sort of local fluctuation is readily included in the
pseudospin model introduced in the next section.

The spin gap proximity effect, enhanced by a large $V_c$ and small 
$\tilde v_F$, gives a robust mechanism for the perturbative relevance of
pair hopping for a wide and physically reasonable range of interactions.
Similar conclusions can be drawn from examining the perturbative expression
for the beta function for $t_{sp}$ in powers of the interaction strength,
as is discussed in Appendix A.

\subsubsection{Composite Order Parameter}

In the rest of this paper, we shall use the canonical transformation
(\ref{eq:unitary}) in a slightly different way by taking $\lambda = 1$,
which is similar to the transformations employed\cite{orbitalK,oron} in the
analysis of Kondo and orbital-Kondo arrays in one dimension.
The special values of
the coupling constants $V_c$ and $J_c$ for which the quadratic part of the
charge Hamiltonian ${\cal H}_0^c$ is diagonalized  at the point $\lambda = 1$
are the analog of the Toulouse limit of the Kondo problem and
the various decoupling lines of the multi-channel Kondo problem, and
Kondo lattice problems\cite{orbitalK,oron,EK1,toulouse,Ye}.

For $\lambda = 1$, the transformation eliminates the $\theta_c$ dependence
of $U^{\dagger}_1{\cal H}_{pair}U_1$ since
$U^{\dagger}_1[\tilde \theta_c-\theta_c]U_1 = \tilde\theta_c$.
Remarkably, this also implies that the transformed
$\tilde\theta_c$ is gauge invariant. Consequently it is possible
to define a composite superconducting order parameter\cite{EK1,odd} in terms of $U_1$ as,
$O_{comp}=U_1\tilde \psi_{1,\uparrow}\tilde \psi_{2,\downarrow}U^{\dagger}_1
=(2\pi a)^{-1}\exp[-i \sqrt{2\pi}(\tilde\theta_c-\theta_c + i \tilde \phi_s)]$,
which can exhibit long-range
order at zero temperature, despite the constraints of the Hohenberg-Coleman-
Mermin-Wagner theorem for a conventional
order parameter.   Indeed, as discussed in Appendix B, long-range
composite order implies a broken Z(2) symmetry, which, for lack of a better
name, we call $\tau$ symmetry.

The transformation introduces a $\tilde \phi_c$ dependence into the $g_3$
term of ${\cal H}^2_c$, which complicates the analysis somewhat although,
as we shall see, it can be handled. However, whenever $g_3$ can be neglected,
the unitary transformation completely decouples the charge modes of the 1DEG
from the environment. This already constitutes a partial solution of the
problem. Moreover the results are generic for all values of the
couplings in the basin of attraction of the paired-spin-liquid fixed point because,
as we shall show, $\Delta V_c$ and $\Delta J_c$ are perturbatively irrelevant.

\subsubsection{Transformation to Holon Variables}

Having separated spin and charge, it is useful for many purposes to express
the charge excitations as spinless fermions, which we shall call ``holons''.
For the environment Hamililtonian this is accomplished by rescaling the
charge fields of the environment by the real space version of a Bogoliubov
transformation:
\be
\tilde \phi_c \rightarrow {\tilde \phi}_c/\sqrt{2}, \ \ \
\tilde \theta_c \rightarrow \sqrt{2}{\tilde \theta}_c.
\label{eq:rescale}
\ee
which also changes $\tilde K_c \rightarrow 2 \tilde K_c$. Then, using
Eq. (\ref{eq:bosonrep}) for spinless fermions, the Hamiltonian for the
environmental charge excitations can be writen
\ba
{\tilde {\cal H}}_c=&&\tilde v_F \big[{\tilde\psi}_{1,c}^{\dagger} i\partial_x
{\tilde\psi}_{1,c}-{\tilde\psi}_{2,c}^{\dagger} i\partial_x
{\tilde\psi}_{2,c}\big] \nonumber \\
- && \tilde \mu \big[{\tilde\psi}_{1,c}^{\dagger}
{\tilde\psi}_{1,c} +{\tilde\psi}_{2,c}^{\dagger}
{\tilde\psi}_{2,c}\big] + {\tilde g}{\tilde\psi}_{1,c}^{\dagger}
{\tilde\psi}_{2,c}^{\dagger}{\tilde\psi}_{2,c}{\tilde\psi}_{1,c}
\nonumber \\
+  && \frac{g_3} {2 \pi a}
\big[{\tilde\psi}_{1,c}^{\dagger} {\tilde\psi}_{2,c} +H.c. \big]
\label{eq:envcf}
\ea
where $\tilde v_F = v_c (4 \tilde K_c^2 + 1)/ 4 \tilde K_c$ and
$\tilde g = 2 \pi \tilde v_F (\frac {4\tilde K_c^2 -1} {4\tilde K_c^2 +1})$.
The holons, which are created by the operator
${\tilde\psi}_{\lambda,c}^{\dagger}$,
are free fermions at the point $\tilde K_c = 1/2$,
or $\tilde g = 0$.  We can similarly refermionize the charge part of the
pair-tunnelling term to obtain, when $\lambda = 1$,
\ba
U_1^{\dagger}{\cal H}_{pair}U_1
= && \big ({t_{sp} \over \pi a}\big )
\big[{\tilde\psi}_{1,c}^{\dagger}
{\tilde\psi}_{2,c}^{\dagger} + {\rm H.c.}\big]  \nonumber \\
\times && \cos[\sqrt{2\pi}\phi_s] \cos[\sqrt{2\pi}{\tilde\phi_s}].
\label{eq:pairhop*}
\ea
Thus, the pair-tunnelling term couples the holon pair creation operator
in the environment
to the joint spin fluctuations of the 1DEG and the environment.
(In this way, pair-tunnelling can, under the right circumstances,
induce a spin-gap in the environment, even if there is no
preexisting gap.)
Finally, the charge density and current density interactions between
the 1DEG and the environment ($V_c$ and $J_c$) can be written
simply in terms of the usual fermionic expressions for the charge and
current densities, respectively.

A similar transformation to holon variables may be made 
for the charge degrees of freedom of the 1DEG.

\section{The Pseudospin Model}

The general model discussed in the previous two sections cannot be
solved exactly, although it can be studied using the sophisticated
mean-field theory which will be introduced in Section V.
However, the low-energy physics may be extracted from the
solution of any model which has the same degrees of freedom and symmetry as
the original model, and is controlled by the same strong-coupling
fixed point. Here we introduce a ``pseudospin'' model which preserves the
essential physics, yet it is exactly solvable\cite{schannel}.   

The essential point is that the frustration of the motion of holes in an
antiferromagnet\cite{trugman} implies that the interaction between holes in 
the environment is effectively strong, {\it i.e.} $\tilde K_c$ and 
$\tilde v_c$ are small.
Thus we may ignore the bandwidth of pairs of holons in the environment and 
characterize them by a single renormalized excitation energy
$\varepsilon^*$. Then we introduce a pseudospin
operator, $\tau_R^z$, such that $\tau_R^z=+1/2$ if there is a holon
pair in the environment in the neighborhood of $R$, and $\tau_R^z=-1/2$
otherwise.  (Formally, if $\tilde K_c = 1/2$, then it follows
from Eqs. (\ref{eq:envcf}) and (\ref{eq:pairhop})
that the pseudospin raising
operator is given by $\tau^+ = {\tilde\psi}_{1,c}^{\dagger}
{\tilde\psi}_{2,c}^{\dagger}$.) Since the pseudospins are discrete variables,
we must put them on a lattice, where the lattice constant $\xi_p$ represents
the distance the holon can diffuse in an imaginary time $1/\varepsilon^*$.
($\xi_p \sim \sqrt{\tilde v_c^2/\tilde \Delta_c\varepsilon^*}$.)
Evidently, the lattice spacing is the residual effect of the holon
bandwidth in the environment.

The (transformed) Hamiltonian can be expressed in terms of the pseudospins as
\ba
U^{\dagger}_1H_{pseudo}U_1 = && H_{1DEG} +\tilde H_s\\
\label{eq:pseudo}
+ && \sum_R J_{sp}\tau_R^x
\cos[\sqrt{2\pi}\phi_s] \cos[\sqrt{2\pi}{\tilde \phi_s}] \nonumber \\
+ && \sum_R \big \{ \varepsilon^*  - 2\sqrt{2/\pi}
\Delta V_c \partial_x\phi_c \big \}[\tau_R^z + 1/2], \nonumber
\ea
where $H_{1DEG}$ is the Hamiltonian of the 1DEG (with $g_3=0$)
defined in Eq. (\ref{eq:H1DEG}), $\tilde H_s$ is the Hamiltonian for the
environmental spin degrees of freedom, which is the environmental
pieces of $H_s$ defined in Eq. (\ref{eq:hs}),
$U_1$ is defined in Eq. (\ref{eq:unitary}), and for simplicity we have ignored
the term proportional to $\Delta J_c$, which we expect to be small.
The sum is over sites in the pseudospin array,
and it is implicit that the terms involving the continuous
fields are integrated over a cell of size $\xi_p$ about the
site $R$.  We will refer to this simplified model of the
dynamics of the environmental charge degrees of freedom
as the ``pseudospin'' model.

It is important to note that the pseudospin model could have been introduced 
at the outset to represent the active environment, without reference to a
more detailed electronic model. 
In that case, $H_{pseudo}$ could be written in terms of the original 
variables as
\ba
H_{pseudo} = && H_{1DEG}+\tilde H_s \nonumber \\
+ && \sum_R J_{sp} [P^{\dagger}_R\tau_R^+ + {\rm H.c.}]\cos[\sqrt{2\pi}
\tilde \phi_s] \nonumber \\
+ && \sum_R [ \varepsilon  + 2V_c \rho_R ] [\tau_R^z + 1/2],
\label{eq:pseudo2}
\ea
where
\be
P_R^{\dagger} = \int_{|s-R|<\xi_p/2} dx P^{\dagger}(x),
\ee
and
\be
\rho_R = \int_{|s-R|<\xi_p/2} dx \rho(x) ,
\ee
are the pair creation and charge density operators defined in Eqs.
(\ref{eq:pairop}) and (\ref{eq:trippair})
respectively, and manifestly $[\tau_R^z+1/2]$ is the holon pair density
operator in the environment.
To see that this is equivalent to the form of the pseudo-spin model
discussed above, we apply the pseudo-spin version of the unitary
transformation, $U_1$,
\be
U=\exp\{-i \sqrt{2\pi} \sum_R \tau_R^z\theta_c\} ,
\label{eq:taushift}
\ee
to Eq. (\ref{eq:pseudo2}).  In this way, we obtain
the transformed version of $H_{pseudo}$ given in
Eq. (\ref{eq:pseudo}) with $\varepsilon^*=\varepsilon - 2V_c$.
It is clear from the derivation that $H_{pseudo}$
has the same symmetry as the starting Hamiltonian.

In the pseudospin model, the umklapp scattering ($g_3$) term of
${\cal H}^2_c$ is unchanged by the transformation $U$, since the
argument of the cosine is displaced by the trivial phase $4 \pi \tau^z_R$,
with $\tau^z_R = \pm 1/2$. Thus, in the pseudospin model, the canonical
transformation decouples the charge degrees of the 1DEG from the environment,
even in the presence of a non-zero $g_3$!

The pseudospin model clearly captures the essential physics of charge
fluctuations in the environment in the limit of small kinetic energy.
In addition it is more general, insofar as it is also
a reasonable representation of the spin gap centers, discussed above.
Of course a continuous distribution of
centers corresponds to the case in which there is an environmental spin gap
everywhere.

\section{Exact Results for the Pseudospin Model with $\varepsilon^*
=0$ at $T=0$}

In this section, we present an exact solution of the pseudospin
model, Eq. (\ref{eq:pseudo2}),
at a suitably chosen decoupling point, in order to elucidate the mechanism
by which a stripe coupled to a magnetic insulating
environment by pair hopping develops a gap in its spin excitation spectrum.
We treat both the case in which there is a preexisting environmental
spin gap and the case in which the environmental spin excitation spectrum
is gapless.  In both cases, the ground state of the
solvable model is a fully gapped paired-spin-liquid state.  However, we
consider the former case to be the more physically relevant, as without
a preexisting environmental spin gap, it is less likely that the model
with physically reasonable values of the bare interactions will lie in
the basin of attraction of the paired spin liquid fixed point. 
A gapped spin liquid 
is the one-dimensional version of singlet superconducting pairing, although 
it also displays enhanced charge density wave correlations\cite{Boso1D,ZKL}

\subsection{The decoupling limit}

The close formal relation between the pseudospin model $H_{pseudo}$ and
a Kondo lattice suggests that there is a counterpart of the solvable limits
of the one dimensional Kondo\cite{oron} and orbital Kondo\cite{orbitalK}
arrays that
we have analyzed previously.  This is in fact the ``decoupling limit'',
discussed earlier, in which
$\Delta V_c = 0$ ({\it i.e.} $V_c=\pi v_c/2K_c$), so that the
unitary transformation, $U$, decouples the
charge degrees of freedom of the 1DEG from the remaining degrees of freedom.
The spin part of the Hamiltonian remains nonlinear and, in general,
it involves the dynamics of the pseudospins.  However, a further
great simplification occurs in the
limit $\varepsilon^* \rightarrow 0$ ({\it i.e.} $\varepsilon=2V_c$)
at which point the pseudospin operators, $\tau^x_R$,
commute with the transformed Hamiltonian, $U^{\dagger}H_{pseudo}U$,
so the set of eigenvalues, $\tau_R^x=\pm 1/2$, are good quantum numbers.

In the ground state, the {\it transformed}
pseudospins are ordered, {\it i.e.} $\tau^x_R =\tau_{0}^x$
for all $R$, and there is a two-fold degeneracy,
corresponding to $\tau_0^x=\pm 1/2$.
This does not correspond to long-range superconducting order,
(which is forbidden in one dimension)
even though the untransformed $\tau^+_R$ creates charge 2.
After the unitary transformation in Eq. (\ref{eq:shift}), $\tau_R^x$ becomes
the gauge-invariant order parameter which characterizes the
{\it composite} pairing of the holons, and it cannot be expressed as a local
function of the original physical fields.  A similar composite ordering was
discovered for the two-channel Kondo problem\cite{EK1}.
Here the only symmetry that is broken in
the ground-state is the discrete ``$\tau$'' symmetry, discussed
in Appendix B.

We show below that, so long as $J_{sp} \ll W$,
the array of pseudospins is so dense that its discreteness may be ignored in
the ground state\cite{dense}. Then the spin fields are governed by the double
sine-Gordon Hamiltonian
\ba
{\cal H}_s = & & {\cal H}_s^0 + {\cal H}_s^2  \nonumber \\
+ & & \frac {J_{sp}} {2 \pi a} \int_{-\infty}^{\infty} dx
\nonumber \\
\times & & \cos \big [\sqrt{2\pi} \phi_s(x) \big ].
\cos \big [\sqrt{2\pi} \tilde \phi_s(x) \big ].
\label{eq:Hs}
\ea
where 
${\cal H}_s^0$ and ${\cal H}_s^2$ are given in Eqs. (\ref{eq:spin0}) and
(\ref{eq:spin2}), respectively.  We can obtain exact solutions of the spin
part of problem in two different limits.

\subsubsection{The case of an environment with a large spin gap }

We first consider the case in which there is a preexisting spin gap in the
environment and show how it is communicated to the 1DEG.  In terms of our
model,  this corresponds to the case in which ${\tilde K}_s <1$ and
$|{\tilde g}_1/{\tilde v}_s|$ is large. Then the term proportional to
$\tilde g_1$ is relevant (in the renormalization group sense) and
even in the absence of coupling to the 1DEG
produces  a spin gap ${\tilde \Delta}_s$ in the environment.
At energies and temperatures small compared
to ${\tilde \Delta}_s$, the fluctuations of $\tilde\phi_s$ are
effectively pinned, and $\cos \big [\sqrt{2\pi} \tilde \phi_s(x) \big ]$
in Eq. (\ref{eq:Hs})
may be replaced by its expectation value.  Thus, for large environmental
spin-gap, we can readily integrate out the environmental spin-degrees
of freedom, leaving us which a simplified pseudo-spin model
in which  the environmental spin-degrees of freedom no longer appear,
but in which a new effective coupling constant
\be
{\cal J}_{sp}\equiv J_{sp} <\cos[\sqrt{2\pi}\tilde \phi_s]>.
\label{eq:calJ}
\ee
replaces $J_{sp} \cos[\sqrt{2\pi}\tilde \phi_s]$ in the pseudo-spin
Hamiltonian, 
Eq. (\ref{eq:pseudo2}), where $<\cos[\sqrt{2\pi}\tilde \phi_s]>$ is the
zero temperature expectation value.  (This expectation value can be computed
exactly in the continuum limit, $<\cos[\sqrt{2\pi}\tilde \phi_s]>
\sim \tilde \Delta_s/\tilde W$, from known results for the sine-Gordon
field theory, as  discussed below; in the strong coupling limit,
$\tilde \Delta_s\sim \tilde W$, $<\cos[\sqrt{2\pi}\tilde \phi_s]>
\sim 1$.)

Once this replacement is made,
the analysis of this equation is simplified by the fact that the $g_1$
contribution to $H_1^s$ is irrelevant, provided $g_1$ is not too large:
On the one hand, with respect to the non-interacting fixed point
defined by ${\cal H}_0^s$,  the final (pair-tunnelling) term in
Eq. (\ref{eq:Hs}) is perturbatively relevant, while the $g_1$ term is
perturbatively irrelevant.  More to the point, the term
proportional to $J_{sp}$ is a relevant
perturbation relative to the full sine-Gordon Hamiltonian
${\cal H}_0^s + {\cal H}_2^s$, whereas if we  reverse the logic,
and we treat the $g_1$ term as a
perturbation, we find that it is irrelevant.
We therefore drop the $g_1$
term for the present with the result that
${\cal H}_s$ is reduced to a (solvable) sine-Gordon Hamiltonian
for the field $\phi_s$.  As discussed below, the solution of this
problem is qualitatively described by the classical limit,
in that $\phi_s$ is thus pinned in the
ground state, and there is a corresponding spin gap.

\subsubsection{The case of small, bare environmental spin gap}

When the environment does not have a large, preexisting spin gap, we may
omit ${\cal H}_s^2$ in Eq. (\ref{eq:Hs}), and rewrite ${\cal H}_s$ as
\ba
{\cal H}_s = && {\cal H}_s^0 \nonumber \\
+ && \frac {J_{sp}} {4 \pi  a} \int_{-\infty}^{\infty} dx \nonumber \\
\times && \big\{ \cos \big [\sqrt{4\pi} \phi_s^+(x) \big ] +
\cos \big [\sqrt{4\pi} \phi_s^-(x) \big ] \big \},
\ea
where $\phi_s^{\pm} = (\phi_s \pm \tilde \phi_s) /\sqrt 2$. Then, in the
special case in which the spin Hamiltonians of the stripe and the environment
are symmetric ($\tilde K_s = K_s$ and $\tilde v_s = v_s$), ${\cal H}_s$
may be written as a sum of two independent sine-Gordon Hamiltonians in the
variables $\phi_s^{\pm}$. The major difference from the case in which the
environment has a spin gap is that $K_s$ is replaced by $2K_s$.

\subsection{Sine-Gordon models}

Until now, we have considered in parallel the cases in which the environment
has and does not have a preexisting spin gap.  To streamline the subsequent
discussion we will focus solely on the more physically
interesting case in which there is a large preexisting environmental spin gap;
the other case can be straightforwardly analyzed along similar lines.
So, for example, the double sine-Gordon model in Eq. (\ref{eq:Hs})
will be replaced by the ordinary sine Gordon model
in which
${\cal J}_{sp}$ replaces $J_{sp} \cos[\sqrt{2\pi}\tilde \phi_s]$.

The solution of the resulting sine-Gordon Hamiltonians is well 
known\cite{dhn}. The excitations are massive solitons
(which correspond to a ``magnon'' with spin 1 and charge 0)
 with energy spectrum given by
\be
E_s(k)=\pm \sqrt{(v_s k)^2 + \bar\Delta_s^2},
\ee
where
\begin{equation}
\bar\Delta_s \sim
\frac {v_s} a \left[ \frac{{\cal J}_{sp} a} {v_s } \right]^{\alpha},
\label{eq:Ds0}
\end{equation}
with
\be
\alpha=2 /(4 - K_s),
\ee
provided $K_s < 4$. 
In addition, so long as $\alpha <1 $, there are breather modes\cite{dhn},
{\it i.e.} two magnon bound states, with spin zero and energy 
$\sim \bar\Delta_s$. In particular, as discussed in Eq. (\ref{eq:ignore2}), 
spin rotation invariance implies that at low energies, $K_s\approx 1$, which 
in the case of a large environmental spin gap implies $\alpha=2/3$.
For $\alpha=2/3$ there are two breathers, one with energy $\bar\Delta_s$ and 
the other with energy $\sqrt{3} \bar\Delta_s$. The spin gap 
$\bar\Delta_s$ also defines a correlation length, $\xi_s=v_s/\bar\Delta_s$, 
which characterizes the response of the spin field to external perturbations.  
Clearly, it is consistent to ignore the discreteness of the pseudospin array 
so long as $\xi_s \gg \xi_p$.

There are two other classes of excitation of the spin degrees of
freedom, both of which are non-propagating in the decoupling limit,
but which acquire a finite (but large) mass when perturbations are
included.  The first involves a kink in the pseudospin order, so that, for
instance, $\tau^x_R = 1/2$ for $R< 0$ and $\tau^x_R = -1/2$ for
$R\ge 0$.  This induces a corresponding ``half'' soliton in the
$\phi_s$ field, and so corresponds to a ``spinon'' with
charge 0 and spin 1/2 with a creation energy,
\be
\bar\Delta_{spinon} \sim \bar\Delta_s;
\ee
it is unclear at present whether $2\bar\Delta_{spinon}$ is greater than
or less then $\bar\Delta_s$, which ultimately determines whether the
magnon is stable or subject to decay into two spinons.  (Classically,
{\it i.e.} in the $K_s \rightarrow 0$ limit, $2\bar\Delta_{spinon}=\sqrt{2}
\bar\Delta_s > \bar\Delta_s$.)
The second excitation involves a flip of the pseudospin
at one point\cite{tsvelik}.
Again, because the spin $\phi_s$ fields are quite rigid ({\it i.e.} $\xi_s$
is large),
they will hardly respond to such a flip,
so the energy of this
excitation can be estimated as
\ba
\bar\delta= && ({\cal J}_{sp}/\pi a)<\cos(\sqrt{2\pi}\phi_s)> \nonumber \\
\approx && \rho(E_F)\bar\Delta_s^2.
\label{eq:deltabar}
\ea
(The fact that this excitation involves minimal relaxation of $\phi_s$ can
also be seen, {\it a posteriori}, from the fact that $
\bar\delta \ll \Delta_s$.)

\subsection{Correlation Functions}

Since a continuous symmetry  cannot be broken in one dimension,
the ``state'' of the system is characterized by the correlation functions of
the various possible order parameter fields,

In the case of noninteracting electrons, density-density
correlation functions decay as $1/x^2$. Therefore, any correlation function
$C_i(x,x') =<O_i(x)O_i(x')>$ which decays as
$x^{- \alpha_i}$ is ``enhanced'' if $\alpha_i < 2$;  the corresponding
susceptibility diverges  as $T^{\alpha-2}$ in the limit
$T\rightarrow 0$.
The order parameters
whose correlation functions are enhanced are: $2 k_F$ charge density wave
\be
O_{CDW}=[\psi_{2,\uparrow}^{\dagger} \psi_{1,\uparrow} +
\psi_{2,\downarrow}^{\dagger} \psi_{1,\downarrow}],
\ee
and singlet pairing
\be
O_{SP} \equiv P^{\dagger}(x),
\ee
where $ P^{\dagger}$ is defined in Eq. (\ref{eq:pairop}).
At temperatures small compared to the spin gap, $\Delta_s$,
the spin field is massive so the spin fluctuations contribute a multiplicative
constant to these correlation functions, while all others exhibit exponential
decay. Away from half filling, there is a band of solitons and
the exponents are given by $\alpha_{CDW} = K_c^*$ and $\alpha_{SP} = 1/K_c^*$.
Here, $K_c^*$ is the value of $K_c$, renormalized by umklapp scattering.

For $1/2 < K_c < 1$, both SP and CDW correlations are enhanced, but the CDW
correlations decay more slowly with $x$. However, as usual for quasi
one-dimensional systems, disorder and the coupling between stripes determine
the fate of an array of stripes.

Even at zero temperature,
the  correlation function of the untransformed pseudospin operators decays
rapidly with distance.  However, the transformed
pseudospins
$<U\tau^x_R  \tau^x_{R'}U^{\dagger}>$ exhibit long-range order
at $T=0$, and Ising-like behavior at finite temperature,
\be
<U^{\dagger} \tau^x_R \tau^x_{R'} U>
\approx  (m_{\tau})^2 \exp{[-|R-R'|/\xi_{\tau}(T)]},
\label{eq:lro}
\ee
where the temperature dependent values of
$\xi_{\tau}(T)$, which diverges as $T\rightarrow 0$,
and $m_{\tau}$, which approaches $1/2$, are estimated below.
As in the case of the quantum Hall effect\cite{Hall}, integer spin chains
\cite{ga}, and various Kondo arrays \cite{orbitalK,oron},  so also
in the present case the coherent state
of the system is characterized by the long-range order
of a non-local order parameter.

\section{Approximate results for the pseudo-spin model at $T \ge 0$}

Our purpose in this section is to obtain a more complete (but approximate)
solution of the model at finite temperature and finite $\varepsilon^*$. We
will also discuss,
qualitatively, the perturbative effects of deviations from the decoupling
limit of the model ({\it i.e.} the effects of non-zero $\Delta V_c$).
Again, for simplicity, we restrict our attention to the more physically
interesting case in which there is a large preexisting environmental spin gap;
the other case can be straightforwardly analyzed along similar lines.
Recall that in this case, the environmental spin
degrees of freedom can be integrated out leaving us with the
pseudo-spin Hamiltonian, Eq. (\ref{eq:pseudo2}),
with the effective coupling
${\cal J}_{sp}$, defined in  Eq. (\ref{eq:calJ}), 
replacing $J_{sp} \cos[\sqrt{2\pi}\tilde \phi_s]$.

(It is also important to remark that the general model considered previously
can be treated at the same level of approximation.  The results differ
little from those we obtain, here, for the pseudospin model,
which substantiates our view that there is little physically important
difference between the two models.  However,
we have been unable to obtain analogs of the exact results discussed
in the previous section for the general model.)

We have shown in the previous section
that the transformed pseudospins are condensed at $T=0$. 
The important thermal fluctuations which destroy this order are the 
spinon excitations which produce kinks in the order parameter field, as 
discussed above.
Thus, the transformed pseudospin correlation functions at low temperature
are equivalent to those of a classical Ising model with exchange coupling,
$\Delta_{spinon}$.  As a consequence, for sufficiently small $T$,
the correlation length diverges as
\be
\xi_{\tau} \approx \xi_p \exp[\Delta_{spinon}/T].
\label{eq:xitau}
\ee

At first blush, Eq. (\ref{eq:xitau}) might be expected to apply so long as
$T \ll \Delta_{spinon}$, but in fact it only holds so long as
$T \ll \delta$;  this is because at temperatures of order $\delta$,
the large density of thermally excited single pseudospin flips
(which, by themselves, directly affect only the magnitude, but not the
range of the pseudospin order) leads to a large renormalization of the
spinon creation energy;  Eq. (\ref{eq:xitau}) remains valid, but with
a temperature dependent renormalized spinon creation energy replacing
$\Delta_{spinon}$ (and latice constant, $\xi_p$).

We obtain an {\it estimate} of this renormalization using the
technique of Coleman, Georges, and Tsvelik\cite{tsvelik}.
Basically, this amounts to
making a mean-field like decomposition of the non-linear term
({\it i.e.} the term proportional to ${\cal J}_{sp}$) in
$\tilde H_{pseudo}$,
so that in computing the thermodynamic properties of $\phi_s$, we replace
the transformed pseudospin operators $\tau_R^x$ by their thermal expectation
value,
$m_{\tau}= <U^{\dagger}\tau_R^x U>$, and conversely
in computing the pseudospin properties, we treat $<\cos[\sqrt{2\pi}\phi_s]>$
as a pseudo magnetic field.  As with all mean-field theories in 
one dimension, this approximation has the fault that it produces
spurious long-range order at finite temperature, where
$<U^{\dagger}\tau_R^xU>$ and $<\cos[\sqrt{2\pi}\phi_s]>$ are actually 
equal to zero. However, we shall see that the mean-field theory
is exact in the limit $\varepsilon^*$ and $T$ tend to zero, and thus its 
results are reliable at low temperatures when it is used 
to estimate {\it local} quantities such as $\Delta_s$, $\Delta_{spinon}$ and 
$m_{\tau}$. In other words, it is correct for intermediate scale
fluctuations.  (For example, $m_{\tau}$ should be defined in terms of the
asymptotic form of the composite order parameter correlation function
in Eq. (\ref{eq:lro}), and the mean-field theory should be viewed as a
way of estimating it as the ``local'' expectation value of an operator.)

In the mean-field approximation the self-consistent equations for the 
temperature-dependent gaps, $\Delta_s(T)$ and $\delta(T)$ are:
\ba
\delta(T) =&&(2{\cal J}_{sp}/\pi )<cos[\sqrt{2\pi}\phi_s]>,
\label{eq:first}
\\
\Delta_s(T) = &&\bar\Delta_s [2<\tau_R^x>]^{\alpha}, \\
<U^{\dagger}\tau_R^xU>=&&(\delta/4E_b)\tanh[\beta E_b(T)],
\label{eq:tautran}\\
E_b(T) = && \sqrt{(\epsilon^*)^2+(\delta/2)^2},
\label{eq:Eb}
\ea
where $\bar\Delta_s$ and $\bar\delta$ are,
respectively, the values of $\Delta_s$ and $\delta$
at $T=0$ and $\varepsilon^*=0$, as given in Eqs. (\ref{eq:Ds0}) and
(\ref{eq:deltabar}) above.  Finally, $<cos[\sqrt{2\pi}\phi_s]>$ should
be computed at finite temperatures using
known results from the thermal Bethe ansatz\cite{thermBethe}
for the sine-Gordon model.  These
results are quite complicated, but fortunately the information we need is
fairly minimal;  specifically, that $<cos[\sqrt{2\pi}\phi_s]>$ is
a monotonically decreasing function of temperature, with
the scale for the temperature dependence set by the zero-temperature gap.
Among other things, this implies that so long as
$\Delta_s(T)\gg T$, we can use the zero-temperature result
\be
<cos[\sqrt{2\pi}\phi_s]>
\approx (\pi a{\bar\delta}/2{\cal J}_{sp})
[2<U^{\dagger}\tau_R^xU>]^{(2\alpha-1)}.
\label{eq:cosat0}
\ee
for the sine-Gordon part of the calculation.
It is clear from these equations that, for $T \ll E_b(0)$ and
$T\ll\Delta_s(0)$,
all gap parameters are well approximated by their zero temperature values.
Conversely, the gaps begin to decrease
when for $T \sim E_b(0)$ if $E_b(0) < \Delta_s(0)$,
or when $T\sim\Delta_s(0)$ if
$\Delta_s(0) < E_b(0)$. We can, in general, define a characteristic
crossover temperature, $T_{pair}$, to be that temperature at which
$\Delta_s(T)$ begins to drop significantly from its zero-temperature
value.  In some cases, this is the only obvious crossover temperature
in the problem.  However, we will see that under some circumstances,
it is still true that $\Delta_s(T) \gg T$ for a substantial range
of temperatures above $T_{pair}$;  in these cases there is a second,
 parametrically larger crossover temperature, $T_{pair}^{\prime}\gg T_{pair}$,
at which the spin gap gets to be comparable to $T$.  For
temperatures above $T_{pair}^{\prime}$, all effects of 
pairing and coherence are
negligible. 

\begin{figure}
\epsfysize=3.5in
\epsfbox{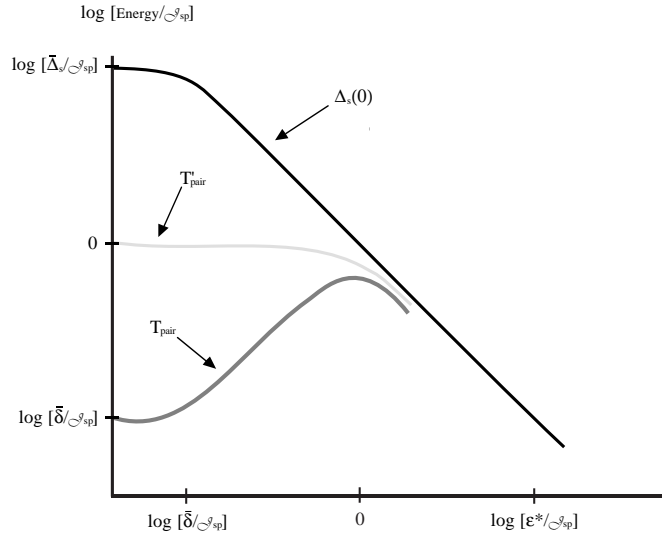}
\caption{Energy scales from the solution of the pseudospin model as
a function of $\varepsilon^*$:  $\bar \delta$ and $\bar \Delta_s$ are,
respectively, the coherence scale and the spin gap derived
from the exact solution of the model for $\varepsilon^*=0$ and given in
Eqs. (48) and (51),
$\Delta_s(0)$ is the zero temperature value
of the spin gap, $T_{pair}$ is the temperature scale at which $\Delta_s(T)$
begins to fall significantly relative to its zero temperature value, and
$T_{pair}^{\prime}$ is the temperature at which $\Delta_s(T)\sim T$.}
\label{fig2}
\end{figure}

We can now proceed to analyze the solution
of these equations as a function of temperature and  
$\varepsilon^*$.  The results
(for the
important case mandated by spin-rotation invariance in which $\alpha=2/3$)
can be sumarized as follows:  $\Delta_s(0)$ is largest for $\varepsilon^*=0$,
and falls slowly, roughly as $\varepsilon^{-1}$, with increasing
$\varepsilon^*$, but only vanishes ({\it i.e.} pairhopping becomes
irrelevant) when $\varepsilon^* \sim [{\cal J}_{sp}]^2/g_1 $.  $T_{pair}$
is much smaller than $\Delta_s(0)$ for small $\varepsilon^*$, but
{\it increases} with increasing $\varepsilon^*$, reaching a maximum
for $\varepsilon^*\sim {\cal J}_{sp}$, at which point all energy
scales are comparable;  $T_{pair}
\sim \Delta_s(0)\sim {\cal J}_{sp}$.  
Meanwhile, $T_{pair}^{\prime}$
is of order ${\cal J}_{sp}$ and roughly independent of
$\varepsilon^*$ for $\varepsilon^*$ small compared to ${\cal J}_{sp}$,
and becomes indistinguishable from $T_{pair}$ for $\varepsilon^* > 
{\cal J}_{sp}$.  These results are shown schematically in Fig. \ref{fig2}.

In the following, we derive these results, focussing sequentially on
four distinct regimes of behavior as a function of $\varepsilon^*$;
in the subsection headings, the ranges are expressed with numerical
exponents for the important case $\alpha=2/3$, as well as
algebraically for general $\alpha$.

\subsection{For the case $\varepsilon^*\ll {\cal J}_{sp}
[{\cal J}_{sp}/W]^{1/3}$,
$\ \  \ \ \ \ \ \ \  \ \ \ \ \ \ \ \ \ \ \ \ \ \
$ {\it i.e.} when $\varepsilon^* \ll \bar\delta$}

In this regime, the results are qualitatively the same as for
$\varepsilon^*=0$.  (Note that for
$\varepsilon^*=T=0$, the self-consistent
equations (\ref{eq:first}-\ref{eq:cosat0}) are exact.)
There is little temperature dependence
of any of the gap parameters in the low temperature regime,
$T\ll T_{pair}$, where
\be
T_{pair}\sim \bar\delta.
\ee
Clearly,
substantial suppression of $\Delta_s(T)$ due to pseudospin fluctuations
begins to occur when $T\sim T_{pair}$;  as a consequence,
$T_{pair}/\Delta_s(0)\sim \rho(E_f)
\Delta_s(0) \ll 1/2$.

There follows an intermediate temperature regime,
$T_{pair}\ll T \ll T_{pair}^{\prime}$, where
\be
T_{pair}^{\prime}\sim {\bar\delta_s}
[{\bar \Delta}/{\bar\delta_s}]^{2(1-\alpha)/(2-\alpha)};
\label{eq:Tpprime}
\ee
in this regime, even though $\Delta_s(T)$ is strongly suppressed,
it is still true that
$\Delta_s(T)\gg T$, so we can approximate
$<\cos[\sqrt{2\pi}\phi_s]> $ by its zero temperature
value Eq.(\ref{eq:cosat0}), with the
consequence that
\be
\Delta_s(T)\approx {\bar\Delta_s}[\beta{\bar\delta}]^{\alpha/2(1-\alpha)}
\label{eq:DsT}
\ee
and
\be
\delta(T)\approx  {\bar\delta}[\beta{\bar\delta} ]^{{(2\alpha-1)}/
{2(1-\alpha)}}.
\label{eq:dbT}
\ee
However, while significant spin pairing
still survives in this temperature range,
the entropy of the pseudospins is recovered, and hence the
specific heat, $C_v \sim [\delta(0)/T]^{1/(1-\alpha)}$, is large.

$T_{pair}^{\prime}$ is the temperature at which  $T =\Delta_s(T)$, where 
$\Delta_s(T)$ is given by Eq. (\ref{eq:DsT}). 
For temperatures $T\gg T_{pair}^{\prime}$,
there is no coherence, no apparent gap in any of the degrees of freedom,
and the problem can be treated using a high temperature
expansion.

We can summarize the heirarchy of scales in this case as
\be
{\bar\Delta_s}\sim\Delta_s(0)\gg T_{pair}^{\prime}\gg T_{pair}\sim
\delta(0)\sim{\bar \delta}\gg\varepsilon^*.
\ee
Specifically, for the $\alpha=2/3$ case,
${\bar\Delta_s}\sim {\cal J}_{sp}^{2/3}$,
$T_{pair}^{\prime}\sim {\cal J}_{sp}$, and ${\bar\delta}\sim
{\cal J}_{sp}^{4/3}$.

\subsection{For the case ${\cal J}_{sp}[{\cal J}_{sp}/W]^{1/3} 
\ll \varepsilon^* \ll
{\cal J}_{sp} \ \ \ \ \ \ \ \ \ \ \ \ \ \ \ \ \ $
{\it i.e.} when
${\bar\delta}\ll \varepsilon^* \ll {\bar \delta_s}
[{\bar\Delta}_s/{\bar \delta_s}]^{2(1-\alpha)/(2-\alpha)}$}

It is easy to see from Eqs. (\ref{eq:tautran})  and
(\ref{eq:Eb}) that larger
values of $\varepsilon^*$ supress the thermal disordering of the
pseudo-spins, and hence removes the anomalous  renormalization of $\Delta_s(T)$
at low temperatures.  At $T=0$, and so long as $\varepsilon^*\gg
{\bar \delta}$,
\be
\delta(0)=\bar\delta [{\bar\delta}/\varepsilon^*]^{1/2(1-\alpha)}
\label{eq:db0}
\ee
and
\be
\Delta_s(0)={\bar\Delta_s}[{\bar\delta}/\varepsilon^*]^{(2\alpha-1)/2(1-
\alpha)}.
\label{eq:Db0}
\ee
If at the same time, $\varepsilon^*\ll T_{pair}^{\prime}$, then $\Delta_s(0)\gg
\varepsilon^*$, so
\be
T_{pair}\sim\varepsilon^*.
\ee
For $T\ll T_{pair}$, there is little temperature dependence of the gaps,
whereas, for $T \gg T_{pair}$, $\varepsilon^*$ falls out of the problem so
$\Delta_s(T)$, $\delta(T)$, and $T_{pair}^{\prime}$ are given
by Eqs. (\ref{eq:DsT}), (\ref{eq:dbT}), and (\ref{eq:Tpprime}), as before.

The remarkable property of this range of parameters is that, as
$\varepsilon^*$ increases, the spin gap at $T=0$ decreases
rapidly (as expected) but the pairing temperature, $T_{pair}$ actually
{\it increases}. In other words, in order to obtain a high temperature
scale for pairing, the charge transfer energy, $\varepsilon^*$,
should be somewhat above
the Fermi energy!  

We can summarize the heirarchy of scales in this case as
\be
{\bar\Delta_s}\gg \Delta_s(0)\gg T_{pair}^{\prime}\gg T_{pair}\sim
\varepsilon^* \gg
{\bar \delta}\gg\delta(0).
\ee
One remarkable feature of this result, which relies on the particular value
$\alpha=2/3$, is that in this regime $\Delta_s(0) \sim [{\cal J}_{sp}]^2/
\varepsilon^*$, $T_{pair}\sim \varepsilon^*$, and 
$T_{pair}^{\prime}\sim {\cal J}_{sp}$ are all
independent of the bandwidth. Note that at the upper end of this range,
$\Delta_s(0)\sim T_{pair}\sim T_{pair}^{\prime}\sim \varepsilon^*\sim
{\cal J}_{sp}$!  This same conclusion follows from evaluating
the expressions in the next subsection at the lower limit of the stated
range.

\subsection{For the case ${\cal J}_{sp} \ll \varepsilon^*
\ll W$, $ \ \ \ \ \ \ \ \ \ \ \ \ \ \ \ \ \ \ \ \ \ \ \ \ \ \ \
$ {\it i.e.} when ${\bar\delta}[{\bar\Delta_s}/{\bar\delta}]^{2(1-\alpha)/
(2-\alpha)}
\ll \varepsilon^* \ll W$}

Whenever ${\bar\delta}[{\bar\Delta_s}/{\bar\delta}]^{2(1-\alpha)/(2-\alpha)}
\ll \varepsilon^*$, it follows that $\Delta_s(0)\ll
\varepsilon^*$.  As a consequence, the temperature dependence of the
various gaps is set by
\be
T_{pair}\sim \Delta_s(0)
\ee
where $\Delta_s(0)$ and $\delta(0)$ are given by Eqs. (\ref{eq:Ds0}) and
(\ref{eq:db0}), above;  moreover, there is no longer a distinct temperature
scale $T_{pair}^{\prime}$.

The heirarchy of scales in this case can be summarized as
\ba
{\bar\Delta_s}&&\gg \Delta_s(0) \sim T_{pair}
\gg{\bar \delta}\gg\delta(0),\nonumber \\
\varepsilon^* &&\gg \Delta_s(0).
\ea
In this regime, both $\Delta_s(0)$ and, correspondingly, $T_{pair}$ are
decreasing functions of $\varepsilon^*$.
To be specific, for the case of $\alpha=2/3$, $T_{pair}\sim {\cal J}_{sp}^2/
\varepsilon^*$ and $\delta(0)\sim T_{pair}\sqrt{\varepsilon^*/W}$.

\subsection{$\varepsilon^* \sim W$: Renormalized interactions}

In the limit of large  $\varepsilon^*$, the dynamical nature of the
collective
mode is unimportant; it could have been integrated out to obtain
new effective interactions in the 1DEG, with retardation and spatial
non-locality limited by the size of $\varepsilon^*$.  Moreover,
since in this limit, holon pairs in the environment exist only as
dilute, virtual excitations,  it is sufficient to compute these
interactions perturbatively in powers of ${\cal J}_{sp}/\varepsilon^*$.
To second order
in ${\cal J}_{sp}$, the Hamiltonian is of the same form as $H_{1DEG}$ in Eq.
(\ref{eq:H1DEG}),
but with a renormalized chemical potential and interactions:
\be
g_1^* = g_1 - \delta g.
\label{eq:deltag}
\ee

\be
K_s^* = K(g_1^*)
\ee

\be
v_s^* = v_s + \delta g /2 \pi,
\ee
where $\delta g = ({\cal J}_{sp})^2/4\varepsilon^*$.

When $g_1$ is small,  $g_1^* <0$ and the pair fluctuations produce a net
attractive interaction in the spin degrees of freedom, which leads to
a spin gap of magnitude\cite{lw}
\be
\Delta_s = 4\sqrt{2\lambda/\pi}(v_s/a)\exp[-1/\lambda]
\ee
where
$\lambda = \rho_s |g_1^*| /a$
and $\rho_s = a/\pi v_s$. It is also clear that there is a corresponding
crossover temperature, $T_{pair} \approx \Delta_s/2
\ll \varepsilon^*$, above which the spin gap vanishes and the spin
excitations are well described as linearly dispersing collective modes
with velocity $v_s^*$.  Again, the charge modes are
completely unaffected by the pairing physics, and so continue to
be described as linearly dispersing modes with velocity $v_c$. Hence
the Drude weight (or, equivalently for the 1DEG, the
zero temperature superfluid phase stiffness ) is unrenormalized.

This analysis is strictly correct only if $\varepsilon^ * >W$, because
it did not take account of retardation, which implies that  the induced 
interaction $\delta g_1$, vanishes for energy
exchange much greater than $\varepsilon^*$. However,
for the physically more interesting case, $W \gg \varepsilon^*
\gg {\cal J}_{sp}$,  the effect
of retardation can be studied using an energy shell renormalization group
scheme, as in the electron-phonon problem\cite{ZKL}.  This improved
treatment produces results that are similar in spirit to those described
above, except that, for energies smaller than $\varepsilon^*$ (when there is
no longer a distinction between the retarded and instantaneous pieces of the
interaction),  the effective interaction has a renormalization, $\delta g_1$,
which is a complicated, but calculable\cite{ZKL}, function of $g_1$,
$({\cal J}_{sp})^2/4\varepsilon^*  $, and $\varepsilon^*/W$.  In all
cases, there is a critical value of the charge transfer
energy $\varepsilon_c\sim({\cal J}_{sp})^2/4 g_1$, such that for larger
$\varepsilon^* > \varepsilon_c$, the renormalized value of $g_1$ is
positive at low energies, and there is no spin gap, whereas 
for $\varepsilon^*<\varepsilon_c$,
$g_1^*$ is negative and a spin gap opens up at zero temperature.
This answers the question of how ``active'' the environment must be.

\subsection{Effects of ``irrelevant'' interactions}

We now consider the effects of various interactions that
we set equal to zero in the decoupling limit.
Because the spectrum of the 
pseudospin model has a gap at the solvable point, all of the omitted
terms are formally irrelevant in the renormalization group sense.  
Of course this does
not give us license to completely ignore these terms;  they
can have large quantitative, and at times qualitative effects on the
physics of interest, even if they do not affect the character of the
true asymptotic behavior of the system.

Let us consider the effects of non-zero $\Delta V$ and,
$\varepsilon^*$ on the nature of the excitations
of the system at zero temperature. When these couplings are small,
their most important qualitative effect is to induce dynamics
for the pseudospins.  In the presence of these terms,
the effective Hamiltonian for the pseudospins, obtained by integrating out
the electronic degrees of freedom\cite{tsvelik}, is qualitatively similar
to (but not precisely equal to)
the spin 1/2 Ising model in a transverse magnetic field,
\ba
H^{eff}\sim && \sum_R[(\bar\delta/2)\tau_R^x + \varepsilon^* \tau_R^z]
\nonumber \\
-&&\sum_{R>R'}[K(R-R')\tau_R^x\tau_{R'}^x + \tilde K(R-R')\tau_R^z\tau_{R'}^z]
\ea
in which $K(R-R') \sim \bar\delta^2/\bar\Delta_s$ and $\tilde K(R-R') \sim
(\Delta V)^2/\bar\Delta_s$ and both have range of order $\xi_s$.
As is well known, a
transverse field induces dynamics (propagation of the kinks) in the spin
1/2 Ising model.

As we have seen, the
other effect of $\varepsilon^*$ is to supress thermal fluctuations
of the pseudospins.  At high temperatures, there is an entropy density
$S=(a/\xi_p)\ln \ 2$ associated with the discrete symmetry of the pseudospins.
For $\varepsilon^*=0$, this entropy is lost at about the temperature
$T_{pair}\sim\bar\delta$, where strong pairing sets in.  In higher dimensional
systems this large entropy is presumably responsible for
heavy-fermion behavior in the model\cite{coleman}; in the
present context it leads to the small ratio of
$T_{pair}/\Delta(0)$.  When $\varepsilon^*>\bar\delta$, the majority
of the entropy associated with the pseudospins is lost at
temperatures greater than $T_{pair}$.  As a consequence, thermal
disordering effects are relatively less severe, and $T_{pair}/
\Delta(0) \sim 1/2$ is rapidly restored.

\section{The Behavior of the Charge Degrees of Freedom:}

We have seen that, in the pseudospin model, the canonical tranformation
decouples the charge degrees of the 1DEG from the environment,
and their fluctuations are described  by the
quadratic Hamiltonian, ${\cal H}_c^0$.  This Hamiltonian describes
a fluctuating superconductor, with phase $\theta_c$, or
in dual language, a fluctuating
charge density wave, with phase $\phi_c$. Evidently, proximity to
commensurability or the existence of an external potential can substantially
modify the physics.

\subsection{The role of Umklapp scattering}

The charge fields of the 1DEG are governed by the Hamiltonian:
\be
\tilde H_c = H_0^c + H_1^c,
\ee
where $H_0^c$ and $H_1^c$ are given in Eqs.(\ref{eq:charge0}) and
(\ref{eq:charge1}).
Now the $c$-number $(4k_F-G)x$ may absorbed into the phase
$\phi_c$, without changing the commutation relations and the quadratic part
of $H_{0}^c$ in Eq. (\ref{eq:charge0}) may be diagonalized by the canonical
transformation $\phi_c \rightarrow \phi_c  K_c^{1/2}$,
$\Pi_c \rightarrow \Pi_c / K_c^{1/2}$. The net result is that the charge
degrees of freedom are described by a sine-Gordon model with a
chemical potential $\mu^*$ given by
\be
\mu^* = {v_c(4k_F - G) \over 4 K_c}.
\label{eq:chem}
\ee
For the strongly incommensurate case, in which $\mu^*$ is large, we
can ignore the umklapp scattering term (proportional to $g_3$);  in
this case the charge excitations are gapless collective modes with
a sound-like dispersion and a velocity, $v_c$, that is unrenormalized
by the interactions with the environment.  Correspondingly, the
Drude weight, or superfluid phase stiffness (which cannot be distinguished in
one dimension in the absence of disorder) are also unrenormalized.

In the nearly commensurate case, which characterizes the doped-insulator
region, the analysis of the corresponding sine-Gordon theory is
the same as for the spin degrees of freedom.
In particular, for
$K_c < 1$, which is always satisfied for repulsive interactions,
the ``particles'' in the theory are massive solitons  with
charge e and spin 0.
It follows at once that the system undergoes an insulator to metal transition
at $|\mu^*| = \Delta_c$, where the chemical potential moves out of the
gap,
and that there is a finite density of solitons:
\be
n_{sol} = {\sqrt{(\mu^*)^2-\Delta_c^2} \over \pi v_c}
\ee
with $\mu^*$ given in Eq. (\ref{eq:chem}). For small $n_{sol}$, the Drude
weight of the stripe is proportional to $n_{sol}$.
This argument is similar to the analysis of the commensurate-incommensurate
transition by Pokrovsky and Talapov \cite{val}, except that they considered
a two-dimensional classical problem, equivalent to the quantum sine-Gordon
problem in imaginary time.

For quarter-filled stripes \cite{half}, $4k_F = 2\tilde k_F = G/2$, so the 
charge density on the stripe and in the environment may jointly lock to the 
lattice. This commensurability effect competes with superconductivity but, 
if the coupling constant is not too large, it may not develop 
beyond the logarithmic temperature dependence that characterizes the early 
stages of renormalization\cite{scalapino}. We are investigating this behavior
as a potential source of the special stability of quarter-filled stripes for 
doping $x < 1/8$ in the {\LCO} family\cite{tranq,yamada}, and the logarithmic
temperature dependence of the resistivity observed \cite{boebinger,tranq} when 
the onset of superconductivity is suppressed.

\subsection{External Periodic Potential}

Here it is assumed that there is an external potential with a wave vector
$q$ which is close to $2k_F$. Then the Hamiltonian must be supplemented
by a contribution
\be
H_e = u \sum_{\sigma} \int_{-\infty}^{\infty} dx
[\psi^{\dagger}_{1,\sigma}\psi_{2,\sigma}e^{i(2k_Fx-qx)} + H.c.],
\label{eq:He}
\ee
which may be written in the boson representation (\ref{eq:bosonrep}) as:

\be
H_e = {2u \over \pi a} \int dx \ \
cos\bigg [\sqrt{2 \pi} \phi_c + (q - 2k_F)x \bigg]
cos \bigg [\sqrt{2 \pi} \phi_s \bigg ].
\label{eq:He2}
\ee
It is straightforward to show that, when the pseudospin representation is
introduced for the charge degrees of freedom of the environment, then
$H_e$ is not changed by the
unitary transformation defined by $U$ in Eq. (\ref{eq:unitary}):
{\it i.e.}  $U^{\dagger}H_eU = H_e$. Moreover, it is clear from the spin
Hamiltonian (\ref{eq:Hs}) that $cos[\sqrt{2 \pi} \phi_s(x)]$
has a finite expectation value so that it may be replaced by a constant in
$H_e$ to obtain the asymptotic behavior of the charge degrees of freedom.
Umklapp scattering may be ignored if it is an irrelevant variable or if
$4k_F$ is sufficiently far from a reciprocal lattice vector.
However, the effect of the periodic potential
is similar to that of umklapp scattering.
The main differences are that the solitons are massive when $K_c < 4$,
(as opposed to $K_c < 1$ for umklapp scattering)
and that $\mu^* = v_c(2k_F - q) / K_c$, which modifies the condition
for the metal-insulator transition.

The physical argument for including such a potential is as follows:
In the ordered state of {\LNSCO}, the holes on a given stripe move in an
effective potential produced by the stripes in a neighboring CuO$_2$ plane.
Since stripes in adjacent planes are perpendicular to each other,
the wave vector of the charge contribution to the effective potential is
given by $q = 2 \epsilon $ in units of $2 \pi /a$, where $a$ is the lattice
spacing \cite{tranq}.
In the same units, $2k_F = n_s /2$, where $n_s$ is the concentration of
doped-holes on a given stripe. The present experimental evidence
\cite{tranq,Tvseps} is consistent with
$\epsilon = 1/8$ and $n_s = 1/2$ and hence $q = 2k_F$ for dopant
concentration $x=1/8$. This is the hole concentration near
which the superconducting $T_c$ is suppressed in the stripe-ordered material
La$_{1.6-x}$Nd$_{0.4}$Sr$_x$CuO$_4$ \cite{arnie1} and in
La$_{2-x}$Ba$_x$CuO$_4$, for which there is indirect evidence of stripe
order \cite{arnie2}.

An array of
stripes will undergo a transition to a superconducting state at a temperature
which is determined by the onset of phase coherence and is proportional to
the superfluid phase stiffness \cite{nature} which, in
turn, is proportional to $n_{sol}$.

In Sec. III we considered the case in which the environmental spin gap
arose because the backscattering term proportional to $\tilde g_1$ was
relevant. For a half-filled band with $\tilde g_3$ also relevant, there is a
broken symmetry ground state with period 2$a$, which produces an external
potential on the stripe, with a wave vector equal to 4$k_F$ when $n_s = 1/2$.
Such a potential is commensurate with the umklapp term $g_3$, so the coupling
between these terms must be taken into account. This is an example in which 
spin gaps with and without a broken symmetry may lead to different 
consequences. The physical case has no broken symmetry.

\section{Spin Gap Center}

Another model of some physical interest has a spin gap at one specific location
as, for example, at an isolated antiferromagnetic region in a metal.
This is an example of a dynamical impurity problem,
in which the conduction electrons couple to a center with internal
degrees of freedom. It is well known that an angular momentum analysis
produces a one-dimensional Hamiltonian involving the radial motion of
incoming and outgoing fermions on the half line $r > 0$, where $r$ is the
distance from the pairing center \cite{toulouse}. Also, it is possible to 
extend the space to all values of $r$ by transforming incoming fermions
for $r > 0$ to incoming fermions at position $-r$. Then the problem is
formally equivalent to a one-dimensional electron gas in which only the
right-going fermions interact with the pairing center.
In the absence of left-going fermions, the operator $P^{\dagger}$, introduced
in Eq. (\ref{eq:pairop}), cannot be defined and only the $\eta$-pairing term
\cite{eta}
\be
P^{\dagger}_{\eta,1}=\psi_{1,\uparrow}^{\dagger}\psi_{1,\downarrow}^{\dagger},
\label{eq:Pback}
\ee
couples to the pairing
center. Triplet pairing terms are omitted because the exclusion principle
requires them to be of the form
$\psi^{\dagger}_{1,\uparrow}\partial\psi^{\dagger}_{1,\uparrow}$
which is less relevant than $P^{\dagger}_{\eta,1}$. (The derivative in the
triplet operator leads to an extra power of $1/x^2$ in the correlation
function.)
Thus a pairing center naturally produces singlet pairing.

We consider the case in which the center has a large spin gap,
so the pseudospin variable (representing charge transfer to the center) 
is the only internal degree of freedom of the center that we retain, 
explicitly. Thus the Hamiltonian is 
\be
H_{center} = H_{1DEG} + H_{\eta}
\ee
where $H_{1DEG}$ is given in Eq. (\ref{eq:H1DEG}), although in the
case in which the metallic degrees of freedom represent a higher
dimensional Fermi liquid, one must set the interactions ($g_a$) to zero.
The bosonized form of $H_{\eta}$ is
\begin{eqnarray}
& H_{\eta} & =  \varepsilon  \tau^z  +
{V \over \sqrt{2} \pi} \tau^z \Phi_c^{\prime}(0) \nonumber \\
& + &\frac {{\cal J}_{\eta}} {\pi a}  \big [ \tau^-
e^{i\sqrt{2}\Phi_{1,c}(0)} +   H. c. \big ] .
\label{eq:BHcenter}
\end{eqnarray}
Here
$\Phi_{1,c}(0) = [ \Phi_{1,\uparrow}(0) + \Phi_{1,\downarrow}(0)]/\sqrt 2$.
In this form the model is equivalent to a single-channel Kondo problem
\cite{toulouse}, and it may be solved by making a unitary transformation
$H_{center} \rightarrow U^{\dagger}H_{center}U$ with
\be
U= exp[- i \lambda \Phi_c(0)\tau^z]
\ee
and choosing $\lambda= \sqrt{2} -1$, for the special point
$V= \sqrt{2} \pi \lambda v_c$.
Then $\tilde H_{center}$ becomes
\begin{eqnarray}
& U^{\dagger} H_{center}U & =  H_{1DEG}+\varepsilon  \tau^z \nonumber \\
& + &\frac {J} {\pi a}  \big [ \tau^-
e^{i \Phi_{1,c}(0)} +   H. c. \big ]
\label{eq:BHtrans}
\end{eqnarray}
This the Hamiltonian may be
``refermionized'' by writing the pseudospin operator in the form
$\tau^+ = \eta d$ where
$\eta$ is an anticommuting $c$-number and $d$ is a fermion annihilation
operator, and inverting the boson representation of fermion fields:
\be
\psi_{c}^{\dagger} =  \eta
{e^{i \Phi_{c} } \over \sqrt{2\pi a}}.
\ee
When written in terms of these variables, the
right-going part of the Hamiltonian becomes
\ba
U^{\dagger} H_{1,center}U = & & -iv_c\int_{-\infty}^{\infty} dx\big[
\psi^{\dagger}_c\partial_x \psi_c ]
\nonumber \\
+ & & {{\cal J}_{\eta}  \over \sqrt{2\pi a}} (d  \psi^{\dagger}_c(0) + H.c.),
\label{eq:toul}
\ea
which is precisely the Toulouse limit from which all of the well-known
behavior of  the single channel Kondo problem may be derived \cite{toulouse}.
This argument strongly suggests that arrays of pairing centers in two and
three dimensions behave like Kondo lattices, and that they should show
heavy-Fermion behavior \cite{coleman}.

Of course a single pairing center in a purely one-dimensional model
should also  exhibit this single-channel Kondo behavior.
This would {\it not} happen if we replaced the pseudospin array in 
Eq. (\ref{eq:pseudo2}) by a single center, because we would have omitted a 
possible $\eta$-pairing interaction, of the form 
${\cal J}_{\eta} \tau_R^+[P_{\eta,1}+P_{\eta,2}]$ in that Hamiltonian.  
While momentum conservation indeed makes this term unimportant for the 
extended array, a spin-gap center, by its very nature, breaks translational 
symmetry and hence permits finite momentum transfer scattering processes.
Including these terms, the total pair coupling at a single spin-gap
center in
Eq. (\ref{eq:pairop}) may be written
\ba
H_{pair}& = & \big\{{\cal J}_{sp}P^{\dagger}(R)
\label{eq:eta} \\
& + & {\cal J}_{\eta}
[P_{\eta,1}(R) e^{2ik_FR}
+P_{\eta,2(R)e^{-2ik_FR}}]\big\} \tau^-_R + {\rm H.c.} .  \nonumber
\ea
If we consider a single center at $R=0$,
and consider the case ${\cal J}_{\eta}=0$,
the left-going fermions at position $x$ may be transformed
to right-going fermions at position $-x$, without changing the Kondo coupling.
Thus the subscripts 1,2 become ``flavor'' labels and we have a two-channel
Kondo problem. However, in this language, the ${\cal J}_{\eta}$
term breaks the ``channel degeneracy'',
and is pertubatively relevant, so it produces a single-channel Kondo problem.
On the other hand the oscillating factors in Eq. (\ref{eq:eta}) make the 
${\cal J}_{\eta}$ perturbatively irrelevant for the array
and moreover, since the mismatch of momenta between the 1DEG and the
antiferromagnet imply that ${\cal J}_{\eta}$ is small
compared to ${\cal J}_{sp}$, the neglect of $\eta$-pairing interactions
for the extended system is justified.  This is analagous to the
behavior found previously for Kondo systems\cite{oron}, where the anisotropic 
single-channel Kondo array, behaves as if it were a two-channel Kondo array,
even though the single impurity version of the model exhibits ordinary 
Kondo behavior.

\section{Discussion}

\subsection{Summary of results}

We have studied a model of a 1DEG in an active environment, focussing
in particular, on the case in which the environment possesses both a
charge gap and a spin gap, and the energy difference between a singlet pair 
of holes in the 1DEG and the environment, $\varepsilon$, is small in
comparison to the bandwidth.  We have discovered a new mechanism for 
producing strong superconducting fluctuations on a high temperature scale, in 
which a spin gap is induced in the regions between the stripes by spatial
confinement, and transferred to the 1DEG by pair tunnelling.
A striking feature of this mechanism of superconductivity,
which may be described as a spin-gap proximity effect,  
is that the pairing ({\it i.e.} the spin gap) is a property  of the
insulating state itself, and it is simply imprinted on the
mobile holes through their virtual excursions into the
insulating regions. We have found that this phenomenon
is robust and, in particular, it survives the presence of strongly repulsive
forward scattering interactions, {\it i.e.} Coulomb repulsion
between electrons. 

We have demonstrated that the physics of this problem is
captured by a simple pseudospin model, for which exact and well-controlled
approximate results can be obtained.  This model includes the most 
important interactions:  the renormalized pair-tunnelling 
matrix element, ${\cal J}_{sp}$ (defined in Eq. (\ref{eq:calJ})),
the renormalized energy cost, $\varepsilon^*$, required to move a singlet 
pair of holes from the 1DEG to the environment, the bandwidth
of the 1DEG, $W\sim E_F$, (which is assumed to be large compared to other
energies), and the exponent, $\alpha$, which characterizes
the spin correlations of the 1DEG. We have used renormalization group 
arguments to show that $\alpha \approx 2/3$ for repulsive, spin-rotationally 
invariant interactions, and we shall use this value of $\alpha$ in discussing 
our results.
		      
We have found that,
generically, this model produces singlet pairing (spin gap behavior) 
at a high temperature, $T_{pair}$:
in the limit $\varepsilon^*\rightarrow 0$,
$T_{pair}\sim {\cal J}_{sp} ({\cal J}_{sp}/W)^{1/3}$, 
while for $\varepsilon^*\gg {\cal J}_{sp}({\cal J}_{sp}/W)^{1/3}$,
$T_{pair}$ is the smaller of $\varepsilon^*$ and 
$\Delta_s(0)\sim {\cal J}_{sp}^2/
\varepsilon^*$.  Remarkably, this means that for small $\varepsilon^*$,
$T_{pair}$ is an increasing function of $\varepsilon^*$, which
reaches a maximum value of $T_{pair}\sim {\cal J}_{sp}$ when
$\varepsilon^*\sim {\cal J}_{sp}$.
Below $T_{pair}$, singlet superconducting and CDW susceptibilities 
diverge as $T \rightarrow 0$, with the stronger divergence typically 
associated with the CDW.  Moreover, this high pairing scale is {\it not} 
accompanied by any significant reduction of the zero temperature
superfluid phase stiffness (Drude weight), {\it i.e.} there is no strong mass 
renormalization. We have also identified a zero-temperature spin gap energy,
$\Delta_s(0)$, which plays the role of the superconducting gap, $\Delta_0$.
In the small $\varepsilon^*$ limit 
the ratio $T_{pair}/\Delta_s(0)\sim \Delta_s(0)/W << 1/2$, while for large
$\varepsilon^*$, $T_{pair}/\Delta_s(0) \approx 1/2$, as in BCS theory.
(The evolution of these energy scales as a function of $\varepsilon^*$
is shown in Fig. \ref{fig2}, and discussed in Sec. VI.) 
The ground state of this model
has a broken, discrete Z(2) symmetry, unrelated to any of the usual space-time
symmetries of the problem, and a corresponding non-local order parameter
which develops a non-zero expectation value in the ground state, and has an
exponentially long correlation length at low temperatures.  
(See discussion of ``$\tau$'' symmetry in Appendix B.) 

\subsection{Interactions Between Stripes and Possible Ordered Phases}

To extend our results to situations in which there is a true phase transition,
we must consider the properties of an array of one-dimensional systems
(stripes).  To avoid misunderstanding, we emphasise
that, for purposes of the present discussion, 
``CDW'' refers to charge ordering {\it along} the stripe direction,
whereas ``stripe order'' implies  charge ordering in the direction 
{\it perpendicular} to the stripes, {\it i.e.} ordering of the stripe 
positions and orientations.  Of course, both types of order are a form of 
generalized charge density wave.

The ultimate nature of the long-range order depends, among other things, on 
the coupling between stripes, which is profoundly
influenced by the intervening antiferromagnetically-correlated regions and,
in particular, by the frustration of hole motion in the antiferromagnet,
which was the driving force for the formation of the stripes themselves.
Thus, this coupling should be smaller than the characteristic energies of the 
electronic correlations along the stripe, considered in this paper. 

With this in mind, the onset of superconductivity in a dilute stripe array 
can be studied by introducing weak interactions between well-separated 
stripes. Single-particle tunnelling
between stripes is an irrelevant perturbation \cite{canadians}, because of the
existence of a spin gap, so we do not expect a crossover to higher-dimensional 
Fermi liquid behavior in this limit. Then the nature of the long-range order 
is determined by pair tunnelling and the Coulomb coupling between stripes.

{\it Effects of Disorder:} 
There are two distinct types of disorder which have very
different effects on the physics of an array of stripes.  The first
is a degree of randomness in the couplings {\it between} stripes, which
may be produced by impurities (as in {\it e.g.} organic conductors) or by
quantum or thermal fluctuations in the stripe configuration.
For a ``self-organized'' quasi one-dimensional system, such as a charged 
stripe array, the latter source of disorder is likely to be the more important.
Disorder of this type favors superconductivity (which
is a $k=0$ order) since it strongly 
frustrates the short-wavelength CDW order associated with the $4k_F$ 
or $2k_F$ instabilities of the 1DEG. This is especially so 
when the stripes are strongly fluctuating.
In the simplest situation, the dynamics of the stripes is slow 
compared to the Josephson plasma frequency, as for example in {\LSCO},
and the disorder is essentially static.  On the
other hand, if the CDW and superconducting fluctuations
are on similar time scales, new physics may emerge; an interesting 
possibility is that there exists a novel quantum critical
point which controls the physics in some region of temperatures and
dopant concentration \cite{castellani}.

The second type of disorder affects the coherence of electronic motion
along a single stripe.  For a single stripe, the back scattering of holes 
from an impurity is always pertubatively relevant for the range of interactions
considered here, because CDW correlations are enhanced\cite{LPimp}. 
However, the localization can be superseded by sufficiently strong
Josephson coupling (pair-tunnelling) between stripes, and there will be
an insulator to superconductor transition 
as the concentration of stripes grows or the Josephson coupling
between stripes is, in any other way, increased, with fixed disorder.
This is in agreement with the evolution of the ground state observed in 
{\LSCO} as a function of doping \cite{yamada}, or applied magnetic 
field\cite{boebinger}.

{\it Symmetry of the order parameter:} If stripe order breaks the four-fold 
rotational symmetry of the crystal, the superconducting order will have
\cite{predict,pokrov2} 
strongly mixed extended-$s$ and $d_{x^2-y^2}$ symmetry!  This
will happen in a stripe-ordered phase, such as in {\LNSCO}, or in a possible 
``stripe nematic'' phase, in which the stripe positional order is destroyed 
by quantum or thermal 
melting or quenched disorder, but the stripe orientational order is preserved.
(Such  phases also would be characterized by large induced asymmetries in the
electronic response in the $ab$-plane. Below we discuss some preliminary 
evidence for a transition to a stripe nematic phase in overdoped {\YBCO}.)

On the other hand, when
the stripes are disordered at long length scales, the thermodynamic distinction
between $s$-wave and $d$-wave superconducting order is well defined;  however, 
even here, if there is substantial orientational order to the stripe
fluctuations at intermediate length scales, the interplay between the two
types of superconducting order is likely to be more complicated and more 
subtle than in conventional, homogeneous materials. For example, one can 
imagine that, even in a phase which is globally $d$-wave, substantial
mixtures of $s$ and $d$-wave order could occur
over mesoscopic scales near surfaces\cite{bachall}
or twin boundaries.  

{\it Superconducting fluctuations:}
A necessary corollary of the stripe model is that, in lightly doped
materials, the temperature scale, 
$T_{pair}$, at which pairing occurs (on a single stripe) is parametrically
larger than the superconducting transition temperature, $T_c$, which is
governed by the Josephson coupling between stripes.  Moreover, 
since the pairing force derives from the {\it local} 
antiferromagnetic correlations
in the regions between stripes, both $T_{pair}$ and $T_c$ must be less than the
temperature scale, $T_{AF}$, below which local antiferromagnetic correlations
develop. A sequence of crossovers is, indeed, observed experimentally in
underdoped high temperature superconductors, and they have tentatively been 
identified\cite{batloggemery} with these two phenomena;  see 
Fig. \ref{fig1}, above, and the discussion below.

\subsection{Phase diagram of the high temperature superconductors}

The schematic phase diagram shown in Fig. \ref{fig1}, shows the global framework
in which our model is related to the properties of the high temperature 
superconductors.  The axes in this figure are temperature, $T$, and doping 
concentration, $x$; hatched lines indicate the most important crossover
temperatures, and the solid lines represent phase transitions to the
antiferromagnetically ordered state at very small $x$,
and to the superconducting state at larger $x$.
(In general, there are additional phase transitions and possibly other 
crossovers, but here we wish to focus only on the central physical issues.)  

The upper crossover temperature $T^*_1$ characterizes the
aggregation of charge (holes) into stripes;  as we have shown elsewhere, the 
driving force for this crossover is frustrated phase separation
\cite{ute,spherical,topo}.
Above $T^*_1$ the holes are more or less uniformly distributed, and 
randomly disrupt antiferromagnetic correlations, while below $T^*_1$, the
self-organized stripe array allows local antiferromagnetic correlations to
develop in the hole-free regions of the sample.  At short distances, low
energy spin fluctuations should come from regions with the character of
odd-leg ladders, and be like those of the one-dimensional Heisenberg model
\cite{ladder}, and, indeed,
there is experimental evidence \cite{stripesgo} indicating 
that this is the case in {\LSCO}. As  $x\rightarrow 0$, 
$T^*_1$ approaches the temperature $T_{\chi}$ at which local antiferromagnetic 
correlations develop in the undoped systems\cite{batlogg}.  
Between $T^*_1$ and the superconducting transition temperature $T_c$,
there is a large range of temperatures in which there are significant stripe 
correlations, but coherence between stripes can be largely ignored; this is 
the region of temperatures
addressed by the calculations in this paper.  As the concentration
of holes increases, the separation between stripes eventually becomes
comparable to their width, at which point all information concerning
the Mott insulating state is lost;  for this reason, we have shown
$T^*_1 \rightarrow 0$ at a dopant concentration, $x_{max}$.

We identify the lower crossover temperature $T^*_2$ with $T_{pair}$, the 
temperature at which
pairing (spin gap) behavior emerges within a stripe.  This is also
the temperature below which significant local, quasi one-dimensional
superconducting fluctuations become significant.  For local
probes of the spin and quasiparticle response functions,
the system should appear all but superconducting below this temperature.
Since $T_{pair}$ is more or less a property of a single stripe, we have
shown it as a relatively insensitive  function of $x$, until it is
cut off by $T^*_1$ at larger dopant concentrations.  From this figure,
it is clear that $T_{pair}$ is substantially greater than $T_c$ throughout
the underdoped regime, and possibly even at optimal doping, and only
approaches closely to $T_c$ in the overdoped regime. 
Thus, in underdoped materials, $T_c$ is determined by the superfluid phase stiffness, 
and hence by the Josephson coupling between stripes, rather than by the 
pairing scale. This is consistent with our previous analysis\cite{nature}.

It should be noted that a phase diagram of the same form as that shown
in Fig. \ref{fig1} has been considered, previously, on purely phenomenological
grounds\cite{batloggemery}, with the crossover
temperatures determined as follows:  

\noindent{\bf 1)}  The upper crossover occurred at a characteristic 
temperature deduced by Batlogg and coworkers\cite{batlogg} from 
an analysis of susceptibility and transport properties, and by Loram and
coworkers\cite{loram} from an analysis of thermodynamic data.
We feel that all of these phenomena are broadly consistent with our 
identification of $T^*_1$ with the emergence of stripe and local
antiferromagnetic order.  (It appears that a pseudogap appears in the c-axis 
optical conductivity\cite{homes} at this temperature.
Much of the c-axis optical oscillator strength will be shifted to energies 
higher than $\tilde \Delta_s + \varepsilon^*/2$ as the stripe correlations 
emerge below $T^*_1$.)  If we accept this identification, then 
for moderate doping concentrations, a typical value is $T^*_1 \sim 300K$,
although it depends somewhat on the particular material, and
rather more strongly on the dopant concentration. Indeed, stripe correlations have 
been seen neutron scattering experiments all the way up to 300K, although
the scattering cross section decreases continuously, making it difficult to
identify them unambiguously at high temperatures\cite{stripesgo}.

\noindent{\bf 2)} The lower crossover was identified by Batlogg and Emery
\cite{batloggemery} as the characteristic ``pseudogap'' temperature, deduced 
from the temperature dependence of the Cu NMR $1/T_1T$, which correlates well 
with the emergence of superconducting gap structure in ARPES experiments
\cite{shen,ding}, and a narrowing of the ``Drude-like'' peak in the optical conductivity 
in the $ab$-plane\cite{basov}.  If we accept this
identification then, for moderate doping, $T_{pair}\sim 150K$, again
depending somewhat on the particular material being studied.

\subsection{Relation to experiments}

\subsubsection{Estimates of the model parameters}

To begin with, it is necessary to estimate the values of the important 
interactions which determine the behavior of the model.  
The physics is driven by the local antiferromagnetic correlations between 
spins, so {\it a priori} we expect the interactions, other than those within 
a single stripe, to be some fraction of $J_{AF}$, which in the high 
temperature superconductors is in the range $1000K - 2000K$ \cite{nrev}.  
For similar reasons, the bandwidth in the environment,
$\tilde W$, is expected to be a few times $J_{AF}$;  numerical simulations
for the square lattice lead to the estimate that the hole bandwidth
\cite{holebw} is approximately $2.2J_{AF}$.  On the other hand, a naive 
estimate of the bandwidth $W$ of the 1DEG is given by the bare value, 
$2t\sim 1eV$, although this is certainly
reduced substantially due to virtual (high energy) single-particle excursions
into the environment, {\it i.e.} leakage of the hole wavefunction into the
insulating neighborhood of the stripe.  

More detailed estimates may be obtained from experiment.  
Since $\varepsilon^*/2$ is the binding energy of
a holon in the stripe, we expect that it also determines
the temperature at which stripes begin to lose their integrity, so
we estimate that $\varepsilon^* \sim 2T^*_1$.    Thus, $\varepsilon^*$ is
certainly remarkably small, $\varepsilon^* \sim J_{AF}/2$, but still
large enough that the peculiarities of the small $\varepsilon^*$ limit are
avoided.  Similarly, if we identify $T_{pair}$ with the spin-gap
temperature deduced from NMR, we can approximately invert the relation 
$T_{pair}\sim {\cal J}_{sp}^2/\varepsilon^*$ to obtain an estimate
of ${\cal J}_{sp} \approx \varepsilon^*$, where the exact numerical
relation between these two quantities depends on numerical amplitudes which
we cannot calculate with any great accuracy.  For this range of parameters,
it also follows that $\Delta_s(0)\sim T_{pair}$, consistent with estimates
of the superconducting gap from photoemission experiments.  
Finally, from the magnitude of the pseudogap observed in c-axis optical
response, we estimate that $\tilde \Delta_s \approx \varepsilon^*$.  
This implies that the cuprates lie in the crossover region between
large and small $\varepsilon^*$ (regimes B and C described in Sec. VI),
which is also the region of maximum $T_{pair}$, as shown in
Fig. \ref{fig2}.  We feel that
these values of $\varepsilon^*$, ${\cal J}_{sp}$, and
$\tilde \Delta_s$ are physically reasonable.

\subsubsection{Does local pairing on stripes 
provide a consistent explanation of the pseudogap 
behavior of underdoped cuprates?}

In the above discussion, we interpreted the experimentally
measured pseudogap behavior in underdoped cuprates as superconducting
pairing in a large range of temperatures above $T_c$.  This behavior
was predicted by us\cite{nature} on the basis of a phenomenological 
analysis of the
relation between the superconducting $T_c$ and the measured zero temperature
superfluid phase stiffness ({\it i.e.} the zero temperature
London penetration depth).  It provides a very natural explanation
of the ``spin gap'' behavior that has been widely observed in planar copper
NMR measurements in underdoped cuprates\cite{T1T}.  Here, there is a peak 
in $1/T_1T$ at a characteristic pairing temperature above $T_c$, below which
there is a rapid falloff that is quite similar to that observed below $T_c$
in more heavily doped cuprates.  The interpretation of the spin gap
as a superconducting gap has recently received considerable support from 
ARPES experiments\cite{shen,ding} which find that the magnitude and wave 
vector dependence of the pseudogap above $T_c$ is similar to that of the
gap seen well below $T_c$  in both underdoped and optimally doped materials.
The temperature above which this gap structure becomes unobservable correlates
well with the pairing scale deduced from spin gap measurements.  Measurements
of the in-plane optical response are also highly suggestive of superconducting
pairing above $T_c$ in underdoped cuprates\cite{orenstein,schlesinger,basov}.

This interpretation has been questioned because 
a large fluctuation diamagnetism and conductivity
have not been observed between $T_c$ and $T_{pair}$ \cite{millis}.  
However, we believe that the absence of dramatic magnetic field effects is 
readily understood.  Well above $T_c$, the superconducting
fluctuations are essentially one dimensional, with little effect of
the Josephson coupling between stripes. Consequently, an applied magnetic 
field does not drive any significant  orbital motion until coherence develops 
in two (and ultimately three) dimensional patches, close to $T_c$.
We are currently engaged in more detailed calculations of
these effects, to make this argument more quantitative.  

Recently it has been determined\cite{yamada,arnie1} that in 
underdoped and optimally doped \LSCO, there is
a unique relation between the mean separation between stripes ({\it i.e.} the
half-period of the dynamical incommensurate spin fluctuations)
and the superconducting $T_c$. We have previously predicted such a 
relation\cite{predict} as a natural consequence of the existence of 
superconducting fluctuations on a single stripe and the idea that $T_c$
is determined by the Josephson coupling between stripes.

\subsubsection{Commensurability and Near Commensurability Effects}

The charge density on the stripes (and hence, the value of $k_F$)
is largely determined by the competition between the local tendency
to phase separation and the long-range Coulomb interaction;  however,
there are commensurability effects both within the 1DEG (which tend
to pin $2k_Fa=2\pi/m$ where $m$ is the order of the commensurability)
and transverse to the stripes, which tend to pin the spacing
between stripes at an integer times the lattice constant\cite{topo}.
In {\LSCO}, neutron scattering evidence supports the notion that
there is a strong tendency toward locking the hole density within
a stripe near commensurability $m=4$ for a range of $x$ less than
$x=0.125$, and to pin the spacing between stripes near 4 lattice
constants for $x > 0.125$. (See Sec. VIIB.) Within the theory of the 1DEG, 
commensurability leads to a charge gap and insulating behavior. However,
for a weak commensurability, the gap develops at low temperatures where it
must compete with superconductivity. (For an alternative view, see 
Ref. \cite{NW}.)  

\subsubsection{Are there any experimentally testable predictions that
can be made on the basis of this mechanism?}

To begin with, it is important to stress that there already exists considerable
experimental evidence that the physics discussed in this paper is pertinent
to the high temperature superconductors.  Some of this has been
discussed above.  Neutron scattering and transport measurements provide 
direct evidence of hole-rich metallic stripes in an antiferromagnetic
environment in at least the {\LCO} family of materials.
The convincing experimental evidence that underdoped cuprates behave like 
granular materials in that a superconducting gap opens well above $T_c$, 
strongly suggests that
the superconductivity is inhomogeneous at some intermediate scale of length 
and time.  Moreover, the absence of strong effects of magnetic fields in a 
regime of strong superconducting fluctuations indicates that these 
inhomogeneities are likely to be one-dimensional in character.
The fact that both s-wave and d-wave symmetry are manifest in different
phase-sensitive experiments on essentially the same materials supports the
idea that there are strong, local fluctuations which break the (approximate) 
four-fold rotational symmetry of the crystal\cite{dvss,pokrov2}.

However, while we feel that these experimental facts provide strong evidence 
for the general form of our model, they do not probe
microscopic structure of the proposed pairing mechanism. 
There are, however, various signatures that could, in principle, be detected.  
We predict a spin-1, charge zero excitation (a quasi one-dimensional,
magnon mode) with an energy gap $\Delta_s$, which is of order the
superconducting gap.  This mode could, in principle, be detected in
neutron scattering.  We also predict 
one or more gapped charge=0, spin=0 modes, the breathers;  for the expected
case of $\alpha=2/3$ there are two such modes, and the lowest energy one
should also have energy gap $\Delta_s$.  This mode could, in principle,
be observed by Raman scattering\cite{raman}.  Since it  also could hybridize 
with a
phonon, it could also show up in neutron scattering.  It is interesting to
note that a phonon mode which is sensitive to the onset of superconductivity,
\cite{raman,klein} and a magnon, both with energy about 40meV have been 
observed\cite{keimer} in the superconducting state of {\YBCO};  we are
currently exploring whether these two phenomena reflect the two
collective modes discussed above.

A stripe structure may have a nematic phase, in which the stripes are 
orientationally ordered along a particular direction. Such a phase should
display a striking anisotropy in its phase stiffness. It is interesting to
note that a big increase in the phase stiffness is observed as {\YBCO} is
overdoped\cite{muon}. This behavior has been attributed to superconductivity
(induced by the proximity effect) in the CuO chains, as they become filled. 
However, such an interpretation requires that the superfluid density in the
chains is greater than in the planes, where it originated. Experimentally
it may not be easy to distinguish nematic stripe order in overdoped {\YBCO} 
given the existence of the CuO chains.

One feature of our model is that there are two, physically distinct, 
spin gaps, one associated with the 1DEG, and hence with the ``superconducting
gap'', and the other (larger gap) with the insulating environment.  
However, in practice, we expect that the two gaps will be similar in 
magnitude because the difference will be ``smoothed out'' by the motion of
the holes between the stripe and the environment.  (Exactly this
sort of ``smoothing out''
of the gap occurs in the ``Cooper limit'' for the conventional proximity
effect.)
Finally, we observe that there
are calculable consequences of our model for single particle properties,
such as the density of states, which are currently under investigation.

Another qualitative test of our ideas is to look for {\hty} in new materials 
that have one-dimensional metallic and spin-gapped regions in close
electrical contact built into their structure, 
and not necessarily self-organized. In this regard, we note that
a material with even-leg 
undoped ladders (which have a spin gap\cite{ladder}) in intimate contact with 
doped CuO$_2$ chains should display the mechanism of superconductivity that 
we have proposed here. Interestingly, superconductivity with $T_c = 12K$ has 
been observed \cite{uehara} at a pressure of 3GPa in 
Sr$_{0.4}$Ca$_{13.6}$Cu$_{24}$O$_{41.84}$, a material with this kind of 
structure, although the chains and ladders are in different planes, so the
electrical conatct is not as strong as we would like. At atmospheric pressure, 
it appears that the doped holes 
are in the chains\cite{chainladder} but, at present, it is not known if this 
feature persists at the high pressures required for superconductivity.
		 
Our model also could be studied by numerical techniques. In particular, an
environment with a spin gap could be represented by either a two-leg ladder 
or an incommensurate dimerized half-filled chain. An environment without a 
spin gap would be a half-filled one-dimensional Hubbard model. In either case 
the coupling to the 1DEG should involve strong single-particle or pair hopping 
and a repulsive interaction between holes.

{\bf Acknowledgements:}  We would like to aknowledge important insights
we have gained in discussions with D.~Basov, R.~C.~Dynes, E.~Fradkin,
D-H. Lee, S.~Sondhi, and J.~Tranquada.
SK would like to acknowledge the hospitality of the Physics Department
at UC Berkeley, where this work was initiated, and the support
of the Miller Foundation, through a Miller Visiting Professorship,
and from a John Simon Guggenheim Foundation, Fellowship. 
This work was supported in part by the National Science Foundation grant 
number DMR93-12606 (SK) at UCLA.  Work at
Brookhaven (VE) was supported by the Division of Materials Science,
U. S. Department of Energy under contract No. DE-AC02-76CH00016.

\appendix 
\section{\bf Perturbative Renormalization Group Analysis}

There are three related senses in which we use the renormalization group
to analyze a complex physical problem, such as the present one.  

1)  Firstly,
the renornmalization group, and in particular the notion of fixed points,
is a theory of theories, and
it provides a  context and structure which allows the problem to be
approached in the context of its global phase diagram.
Even when calculations are not carried
out by use of the renormalization group, the  results are fundamentally
informed by its structure.   
For instance,
so long as an exactly solvable model, and a particular problem of physical
interest are governed by the same fixed point, the solvable model can
be said to be an accurate representation of the low-energy physics of the 
problem of physical interest, whether or not there is a microscopic 
correspondence. It is in this sense
that a large class of physically diverse one-dimensional systems can all
be described as ``Luttinger liquids'', or that the resonant level model
represents a solution of the antiferromagnetic Kondo problem.
Similarly the exact solution of the pseudospin model,
presented in Sec. V describes the physics
of the paired spin liquid phase of the 1DEG in an active environment.

2)  The notion of an unstable fixed point (or line of fixed points)
also underlies the use of field theories to describe
condensed matter systems.  Of course, condensed matter systems have
a finite lattice spacing.  However, in the
proximity of an unstable fixed point, 
the correlation length diverges, so that the continuum
limit is actually realized when the correlation length diverges,
but this is equivalent to holding the correlation length fixed, and letting
the bandwidth diverge, as is done in defining a field theory.  Thus, all
the field theory results we employ, incuding the results based on
the equivalence between different field theories which goes under the
title of bosonization, are based on the proximity of the system to the
Luttinger liquid line of unstable fixed points.  

3)  The renormalization
group is also a computational scheme, which in most cases must be
carried out in the context of a perturbative evaluation of the beta
function.  The terms ``relevant'' or ``irrelevant'' in the renormalization group
sense refer to the results of a perturbative evaluation of
the beta function in the neighborhood of a particular fixed point.  
Such methods are useful for determining the
stability or lack thereof of a particular fixed point.  However,
in the case in which there is one or more relevant interaction, these
results can only be used
to guess the nature of the actual ground state.

\subsection{Perturbative treatment of ${\cal H}_{int}$}

One approach to the problem is to treat ${\cal H}_{int}$ as
a small perturbation.  Thus, one imagines determining the properties
of the fixed point corresponding to the decoupled
problems of the 1DEG and the environment, and then assessing the
relevance of ${\cal H}_{int}$ at that fixed point.
Because, by assumption, the environment has a charge gap, 
any interaction involving excitations
of the charge degrees of freedom of the environment is irrelevant
in the renormalization group sense.  Thus, 
${\cal H}_{pair}$ and the 
charge and charge-current interactions 
in ${\cal H}_{int}$ ({\it i.e.} the terms proportional to $V_c$ and $J_c$)
are immediately seen to be irrelevant.
In the case in which the environment has a preexisting spin-gap, the
same analysis implies that the
remaining interactions in ${\cal H}_{int}$ are also perturbatively
irrelevant.  Even in the case in which the environment has gapless
excitations ($\tilde g_1 >0$), the spin couplings can readily seen to
be perturbatively irrelevant.  Thus, for weak enough coupling between
the 1DEG and the environment, the coupling can be ignored in the sense
that the low energy behavior is qualitatively similar to that of the
two subsystems in the absence of their coupling.

In the problem of physical interest, the energy to transfer a pair
of holes from the 1DEG to the environment, $\varepsilon^*$,
is very small compared to the bandwidth.  As
we have shown in the main body of the paper, 
this implies that the perturbative analysis about the
${\cal H}_{int}=0$ fixed point is 
valid only in an extremely restricted regime of parameter space.
In particular, for fixed small, but non-vanishing $t_{sp}$, 
there is a critical value
of $\varepsilon^*$, such that ${\cal H}_{pair}$ is irrelevant for
$\varepsilon^* > \varepsilon_c$, and relevant for 
$\varepsilon^* < \varepsilon_c$. 

\subsection{Perturbative RG about the non-interacting fixed point}

The standard (``g-ology'') treatment of the 1DEG, may be derived by
computing the beta function in powers of the interactions, $g_a$,
using a version of Anderson's poor-man's scaling, in which states at
the band edge are integrated out, and new effective interactions are
computed for the model with a reduced bandwidth, $E < W$.  The 
variation of the coupling constants as a function of $E$ are
determined by a differential equation, in which the microscopoic
values of the interactions serve as initial conditions.  This method
can only be applied if {\it all} the interactions are weak on the scale of
the bandwidth, as it is based on perturbation theory about the non-interacting
fixed point. 
 
For the present problem, one can similarly
derive the appropriate scaling equations
for the entire set of interactions
in perturbation theory about the non-interacting fixed
point.  To do this, we notice that the model defined in Section II
is a particular form of an asymmetric
two-band model, with appropriate couplings, and with bandwidths
$W$ and $\tilde W$, respectively.  However, because of the large difference
in the bandwidths, the integrating
out of high energy degrees of freedom, which is the business end of this
sort of calculation, must be carried out in two stages.
In the
initial stages of renormalization, we integrate out degrees of freedom
(of the 1DEG)
with energies between $W$ and $E$, where $W \ge E \gg \tilde W$.  
The resulting scaling equations apply so long as all the interactions
remain small ({\it i.e.} so long as perturbation theory is adequate) until
$E$ reaches the scale of $\tilde W$.  For further reduction of the bandwidth,
excited states of both the environment and the 1DEG are being simultaneously
eliminated.  In this way, starting with a set of bare coupling constants, one
obtains a set of renormalized coupling constants at the end of the first
stage of renormalization which serve as initial conditions for the second
stage flow equations.

\subsubsection{The RG flows for $\tilde W > E$} 

To begin with, we
ignore the differences in band-
width, so that the model is equivalent to the two-band model
considered by Varma and Zawadowskii \cite{VZ}.  
This allows us to adopt their results
(obtained using the usual methods);  translated into the notation of
the present paper, the scaling equations can be written as
\ba
\dot g_1 &=& -\frac 1 {2\pi \bar v} \big[ 2\alpha g_1^2 + \frac {\beta} 2
\big( t_{sp}^2 - t_{tp}^2 \big)\big] \\
\dot g_c &=& -\frac 1 {2\pi\bar v} \big[ 2\alpha g_3^2 -\frac {\beta} 2
\big( t_{sp}^2 + 3 t_{tp}^2\big)\big] \\
\dot g_3 &=& -\frac {2\alpha} {2\pi \bar v} g_c g_3 \\
\dot U_s &=& -\frac 1 {2\pi \bar v} \big[ \frac {t_{sp}t_{tp}} 2 - 4 U_s^2
\big] \\
\dot U_c &=& \frac 1 {8\pi \bar v} \big[ t_{sp}^2 + t_{tp}^2 \big] \\
\dot t_{tp} &=& \frac 1 {4\pi \bar v} \big[ \alpha ( g_1+g_c) + \beta (
\tilde g_1 + \tilde g_c) - 4U_c - 4U_s \big] t_{tp} \nonumber \\
& & \ \ \ \ \ \ - \frac 1 {\pi \bar v} U_s t_{sp}  \\
\dot t_{sp} &=& -\frac 1 {4\pi \bar v} \big[ \alpha(3g_1-g_c)
+\beta(3\tilde g_1-\tilde g_c) - 4 U_c\big] t_{sp} \nonumber \\
& & \ \ \ \ \ \ \ - \frac {3} {\pi \bar v} U_s t_{tp},
\ea
where $\bar v \equiv (v_F + \tilde v_F)/2$ is the average Fermi
velocity, $\alpha = \bar v/v_F$, $\beta = \bar v/\tilde v_F$,
\ba
U_s\equiv && V_s-J_s \\
U_c\equiv && V_c-J_c 
\ea
and there are three additional scaling equations for $\tilde g_a$ which
can be obtained from the equations for $g_a$ by placing tildes on the
$g_a$'s and interchanging $\alpha$ and $\beta$.  Here, we have augmented
the original equations of Varma and Zawadowskii to include the effects
of umklapp scattering, which was done by Balents and Fisher\cite{BF}.  
(We correct
a factor of two error they made in the scaling equations for $g_3$ and
$\tilde g_3$.)  
Note that we have adopted the opposite sign convention
for the beta function than Varma and Zawadowskii;  here, the dot signifies
the derivative with respect to $\ell \equiv \log[W/E]$, which is the negative 
of their variable, $\log[S]$.

There are several aspects of these equations that are worth noting.  In the
first place, the scaling equation for $t_{sp}$ is the weak-coupling
version of the more general Luttinger liquid result given in Eq. (\ref{eq:alpha});
$t_{sp}$ is perturbatively relevant only if $\big[ \alpha(3g_1-g_c)
+\beta(3\tilde g_1-\tilde g_c) - 4 U_c\big]$ is negative.  
We expect that $g_c$ is negative (but possibly small), 
$\tilde g_c$ is negative and grows in magnitude with renormalization, 
and $g_1$ is positive, but typically decreases with renormalization.
Thus, we see that the two ways in which $t_{sp}$ can become relevant are
through the generation of a large $U_c$, or via spin-gap physics of the
environment, in which case $\tilde g_1$ is negative and grows
with renormalization.  That the latter possibility is the more robust is
further emphasized by the expected large value of $\beta$, which
means that  the term involving $\tilde g_1$ makes the largest
contribution to the beta function.  In either case, by examining the
dependence of the beta functions of the various other interactions
on  $t_{sp}$,
it is clear that once $t_{sp}$ becomes sufficiently large, the
there is a bootstrap effect which accelerates the flows to strong
coupling, in that a large $t_{sp}$ makes a positive contribution to
the beta functions for $g_c$, $\tilde g_c$, and $U_c$, and a 
negative contribution to $g_1$ and $\tilde g_1$.

\subsubsection{The RG flows for $W > E \gg \tilde W$}

We now return to the problem of determining the beta function for
the initial stages of the elimination of high energy degrees of freedom.
The scaling equations for the regime $W\ge E \gg \tilde W$ can be obtained
from the above equations by taking the limit $\tilde v_F\rightarrow \infty$;
this has the effect of projecting out any intermediate states involving
the propagator in the environment.  The result is the scaling equations
which govern the initial renormalization process:
\ba
\dot g_1 &=& -\frac 1 {\pi v_F} g_1^2 \\
\dot g_c &=& -\frac 1 {\pi v_F} g_3^2 \\
\dot g_3 &=& -\frac 1 {\pi v_F} g_cg_3\\
\dot {\tilde g_1} &=& -\frac 1 {4\pi v_F}\big[ t_{sp}^2-t_{tp}^2 \big]\\
\dot {\tilde g_c} &=& \frac 1 {4\pi v_F} \big[ t_{sp}^2 + 3 t_{tp}^2 \big] \\
\dot t_{tp} &=& \frac 1 {4\pi v_F} \big[ g_1+g_c\big]t_{tp} \\
\dot t_{sp} &=& -\frac 1 {4\pi v_F}\big[3g_1-g_c\big] t_{sp} \\
\dot {\tilde g_3}&=&\dot U_s=\dot U_c =0.
\ea

Most importantly from these equations it is clear that, in the initial
stages of renormalization, $t_{sp}$ is {\it reduced} from its
microscopic value, although if the interactions in the 1DEG are not
too strong, this reduction may not be too severe.  There is also an
additive negative contribution to $\tilde g_1$ and a positive
additive contribution to $\tilde g_c$ generated in this initial stage
or renormalization.  This is a form of asymmetric screening which
tends to increase the relevance of $t_{sp}$ in the final stages
of renormalization.  However, it seems to us unlikely that this
latter effect is strong enough to make $t_{sp}$ robustly relevant
at low energies in the absence of an environmental spin gap.

\section{Symmetries of the Model and the Composite Order Parameter}

\subsection{Symmetries of the Model}

To begin with, we tabulate the symmetries of the Hamiltonian of the
1DEG in an active environment, Eqs. (\ref{eq:envham}). 

\begin{itemize}

\item{\bf Parity} is a Z(2) symmetry of the system, which results
in the transformation
\ba
\psi_{1,\sigma}(x)& \rightarrow & \psi_{2,\sigma}(-x), \nonumber \\
\psi_{2,\sigma}(x)& \rightarrow & \psi_{1,\sigma}(-x),
\ea
and the analagous transformation for the environmetal operators.
In terms of bosonic variables,
\ba
\theta_a(x) & \rightarrow & \theta_a(-x) \nonumber \\
\phi_a(x) & \rightarrow & -\phi_a(-x).
\ea
where $a$ denotes $s$ or $c$.
Under the
action of the parity transformation, $P^{\dagger}$, $\rho_c$, and
${\vec\rho}_s$ are even, and $P^{\dagger}_m$, $j_c$, and ${\vec j}_s$ are
odd.

\item{\bf Time reversal} is a second Z(2) symmetry of the system, which results
in the transformation
\ba
\psi_{1,\uparrow}(x) & \rightarrow & i\psi_{2,\downarrow}(x), \nonumber \\
\psi_{2,\uparrow}(x) & \rightarrow & i\psi_{1,\downarrow}(x), \nonumber \\
\psi_{1,\downarrow}(x) & \rightarrow & -i\psi_{2,\uparrow}(x), \nonumber \\
\psi_{2,\downarrow}(x) & \rightarrow & -i\psi_{1,\uparrow}(x), 
\ea
and the analagous transformation for the environmetal operators.
In terms of bosonic variables,
\ba
\theta_c(x) & \rightarrow & - \theta_c(x), \nonumber \\
\theta_s(x) & \rightarrow & \theta_s(x)- \sqrt{\pi / 2}, \nonumber \\
\phi_s(x) & \rightarrow & -\phi_s(x) .
\ea
and, of course, $i\rightarrow -i$.
Under the action 
of the time reversal transformation $\rho_c$ and 
${\vec j}_s$ are even, $P^{\dagger}$, $j_c$, and ${\vec \rho}_s$ are
odd, $P_m^{\dagger}$ transforms as $P_m^{\dagger} \rightarrow
-\exp(i\pi m) P_{-m}^{\dagger}$.
and the corresponding environmental operators transform in the same fashion.

\item{\bf Spin rotational symmetry}  is respected entirely by the model as
originally written, so there is a corresponding SU(2) symmetry of the
system, which transforms the operators according to
\be
\psi_{\lambda,\sigma}  \rightarrow  \sum_{\sigma^{\prime}}
\big [\exp(i {\vec \gamma}\cdot
{\vec\sigma}) \big ]_{\sigma,\sigma^{\prime}}\psi_{\lambda,\sigma^{\prime}}.
\ee
and the analagous transformation for the environmetal operators.
Manifestly, this transformation leaves all the charge, charge current, and
singlet pairing operators invariant, and rotates all spin-vectors in the
appropriate fashion.  Abelian bosonization of the model obscures this
symmetry, which is manifest as a non-trivial relation between $K_s$ and
$g_1$.  Generalizing the original model by defining distinct 
couplings $g_{1,\perp}$ and $g_{1,\parallel}$ would give arbitrary
values of $K_s$ and $g_1$ (which now should be identified with 
$g_{1,\perp}$);  in this case, only the U(1) symmetry associated
with rotations about the z axis remain of the original spin-rotational
symmetry.  The full SU(2) transformation is complicated in terms
of the bosonic variables, but rotations about the z axis correspond to
an additive phase shift to $\theta_s$.

\item{\bf Gauge invariance} or charge conservation, is manifest as a
global U(1) symmetry of the model (since we have not  explicitly included
the gauge fields) which transforms the operators as
\be
\psi_{\lambda,\sigma}\rightarrow \exp(i\gamma)\psi_{\lambda,\sigma},
\ee
and the analagous transformation for the environmetal operators.  
In terms of bosonic variables,
\be
\theta_c  \rightarrow  \theta_c+\sqrt{{2 \over \pi}} \ \gamma, 
\ee
and $\phi_a$ and $\tilde \phi_a$ are invariant.  This
transformation leaves all the particle conserving operators invariant,
and multiplies all pairing operators by a factor of $\exp[-2i\gamma]$.  

\item{\bf Translational (chiral) symmetries:}
There are the two independent symmetries corresponding to 
translations (chiral transformations) of the 1DEG:
\ba
\psi_{1,\sigma} & \rightarrow & \exp(i\gamma_t) \psi_{1,\sigma}, \nonumber \\
\psi_{2,\sigma} & \rightarrow & \exp(-i\gamma_t) \psi_{2,\sigma},
\ea
and the analagous tranformations, defined in terms of a second, independent
angle $\tilde\gamma_t$, for the environmental operators.
In the absence of umklapp scattering, ({\it i.e.} if we set $g_3$ equal zero) 
then $\gamma_t$ can take on any real value between 0 and $2\pi$,
{\it i.e.} there
is an additional U(1) symmetry associated with translations of the 1DEG).
In terms of bosonic variables, 
\be
\phi_c  \rightarrow  \phi_c + \sqrt{{2 \over \pi}} \ \gamma_t 
\ee
and the analagous relations (with $\tilde \gamma_t$) for the environmental
operators.

\item{\bf Spin chiral transformations}

There is an analagous transformation, which amounts to a translation of the
SDW fluctuations by a half a period, in which the up and down spin components
are translated in opposite directions.  We define the spin chiral transformation,
$C$ as
\ba
\psi_{1,\uparrow} & \rightarrow & i \psi_{1,\uparrow}, \nonumber \\
\psi_{2,\uparrow} & \rightarrow & -i\psi_{2,\uparrow}, \nonumber \\
\psi_{1,\downarrow} & \rightarrow & -i \psi_{1,\downarrow}, \nonumber \\
\psi_{2,\downarrow} & \rightarrow & i\psi_{2,\downarrow}, 
\ea
which in
terms of the bosonic variables is,
\be
\phi_s \rightarrow  \phi_s +\sqrt{{\pi \over 2}},
\ee
and we define the analagous transformation for the environmental
operators as $\tilde C$.  $H_{1DEG}$ is invariant under $C$, but it
has the effect of rotating $\vec{\rho}_s$ and $\vec {j}_s$ by $\pi$
about the $\hat z$ axis and changing the sign of 
both $P^{\dagger}$ and $P_0^{\dagger}$, so it is not a symmetry of
the full Hamiltonian;  however $C\tilde C$ manifestly is.
Having said this, it is clear that additional symmetries can be
constructed by combining $C$ and $\tilde C$ with 
spin rotations by $\pi$ about the $\hat z$ axis;  we call these
transformations $R$ and $\tilde R$, and they correspond to
shifts of $\theta_s$ and $\tilde \theta_s$ by $\sqrt{\pi/2}$
respectively.  In this way, an additional discrete group of related symmetry 
transformations  can be constructed consisting of the identity, $C\tilde C$,
$CR$, $\tilde C \tilde R$, $C\tilde R$, and $\tilde C R$;  this group is
Abelian, with a simple group multipliction table, which is readily
obtained.  Notice that, as with time reversal symmetry, this group's
operation on spinor fields is double valued.

\item{\bf $\tau$ symmetry}  There is one additional hidden Z(2) symmetry
of the Hamiltonian, which
combines spin and charge transformations, and which is the symmetry that
is spontaneously broken in the paired-spin-liquid state.  This symmetry is
combines a spin-chiral transformation of the 1DEG, $C$, a $\pi$ rotation
of the environmental spins, $\tilde R$, and an inequivalent gauge
transformation of the charge modes of the 1DEG and the environment.
In terms of the fermionic fields, this symmetry corresponds to the
transformation
\ba
\tilde\psi_{\lambda,\uparrow} & \rightarrow & - \tilde\psi_{\lambda,\uparrow}, 
\nonumber \\
\tilde\psi_{\lambda,\downarrow} & \rightarrow & \tilde\psi_{\lambda,\downarrow}, 
\nonumber \\
\psi_{1,\uparrow} & \rightarrow & i\psi_{1,\uparrow}, \nonumber \\
\psi_{2,\uparrow} & \rightarrow & -i\psi_{2,\uparrow}, \nonumber \\
\psi_{1,\downarrow} & \rightarrow & -i\psi_{1,\downarrow}, \nonumber \\
\psi_{2,\downarrow} & \rightarrow & i\psi_{2,\downarrow}. 
\ea
In terms of bosonic variables, this transformation takes
\ba
\tilde \theta_c & \rightarrow & \tilde \theta_c + \sqrt{{\pi \over 2}} \nonumber \\
\tilde \theta_s & \rightarrow & \tilde \theta_s + \sqrt{{\pi \over 2}} \nonumber \\
\phi_s & \rightarrow & \phi_s  + \sqrt{{\pi \over 2}}.
\ea
This transformation leaves $\rho_c$, $j_c$, $\tilde \rho_c$, and $\tilde j_c$,
invariant, rotates
$\vec\rho_s$, $\vec j_s$, $\vec{\tilde \rho}_s$, and $\vec {\tilde j}_s$
by $\pi$ about the $\hat z$ axis, 
changes the sign of $P^{\dagger}$ and $\tilde P^{\dagger}$,
and transforms $\tilde P_m^{\dagger}\rightarrow -e^{im\pi}\tilde P_m^{\dagger}$ 
and $P_m^{\dagger}\rightarrow -e^{im\pi}P_m^{\dagger}$.

\end{itemize}

In the above, it is important to realize that a shift in the bosonic phases
$\phi_a$ by $\pm \sqrt{\pi/2}$ is equivalent to a displacement through a 
distance equal to the average spacing between the particles.
For $\phi_c$ ($\phi_s$), spins-$\sigma$ are displaced in the same (opposite)
direction. This shift leaves the Hamiltonian of the 1DEG unchanged because 
the arguments of the cosines in the $g_1 \cos(\sqrt{8\pi} \phi_s)$ and 
$g_3 \cos(\sqrt{8\pi} \phi_c)$ terms are changed by $2\pi$. 
To appreciate the significance of this observation, consider the ground-state 
degeneracy of the 1DEG with a half-filled band.  A shift of either $\phi_c$ 
or $\phi_s$ by $\pm \sqrt{\pi/2}$ changes the sign of the operator 
$\psi^{\dagger}_{2,\sigma} \psi_{1,\sigma}$, since its boson representation 
is proportional to $\exp\big[i\sqrt{2\pi} (\phi_c +\sigma \phi_s)\big]$. Thus, if 
this operator is ordered the ground state is two-fold degenerate. This occurs
if both $g_1$ and $g_3$ are relevant, as for example in the negative-$U$ 
Hubbard model with additional nearest neighbor repulsions $V$, and it is
easily understood from a strong-coupling analysis, as the ground state is a
period 2 charge density wave.  These considerations must be taken into
account in studying the full symmetry group of the 1DEG as they imply
that not all the symmetry operations discussed above are linearly independent.

\subsection{The non-local order parameter}

The non-local order parameter defined in terms of the unitary transformation
in Eq. (\ref{eq:shift}) is
\ba
O_{comp} = && U \tilde P^{\dagger} U^{\dagger} \\
= && (\pi a)^{-1}\exp[i\sqrt{2\pi}
(\theta_c-\tilde\theta_c)\cos[\sqrt{2\pi}\tilde\phi_s] \nonumber
\ea
can be expressed as a non-local function of the original fermionic fields
as
\be
O_{comp} = \exp[i \pi \int_{-\infty}^x dy j_c(y) ]\tilde P^{\dagger}.
\ee
Clearly, this composite order parameter is odd under $\tau$ symmetry. 

\section{The Nature of the ``Paired Spin Liquid''}

Various definitions of a ``spin liquid'' are used in the 
literature\cite{spinliq}.
Here, we define a spin liquid to be a quantum disordered ground state of the
spin degrees of freedom of a system, which means that spin-rotation
invariance is unbroken. We also require that translation invariance be 
unbroken for the system to qualify as a liquid. In addition,
to distinguish the spin liquid from a quantum paramagnet and a Fermi liquid, 
we require that a spin liquid support spinon excitations in its excitation 
spectrum.  

The ground state of a spin-1/2 Heisenberg chain is a gapless spin 
liquid\cite{andersonrvb}.  An integer spin chain or a even-leg 
half-integer spin ladder fail to qualify because spinons are confined. 
(The only finite energy states are integer-spin magnons;  spinons
are bound by a linear potential in pairs, or to the
ends of chains\cite{kennedy}.) The frustrated spin-1/2 chain ({\it e.g.} the 
Majumdar-Ghosh model\cite{mg}) fails to qualify because
translational symmetry is spontaneously broken in the ground state.
(See Appendix B.)  The 1DEG away from half-filling displays two kinds
of behavior; a) when $g_{1}$ is irrelevant, it is a gapless spin liquid 
in the universality class of the spin-1/2 Heisenberg chain, 
b) when $g_{1}$ is relevant it has a gap because of spinon pairing and is 
in the universality class of doped polyacetylene\cite{ssh} or a doped 
Majumdar-Ghosh model \cite{KZ}.  It is this latter case, in which spinon
pairing causes a gap or pseudogap in the spinon spectrum, that we call a 
``paired spin liquid'';  spinons are paired in the same way\cite{KRVB} as
electrons in a superconductor, and they must be created in pairs, 
{\it i.e.} by breaking a bound pair which exists in the ``vacuum''.

There are, to the best of our knowledge, only two other theoretically 
well established examples of a spin liquid, according to the above
definition.  The first is the superconducting state of charged particles in 
higher dimensions; in this context, it has been shown\cite{krsc} that the 
usual Bogoliubov quasiparticles have spin 1/2 and charge 0, where both quantum 
numbers are sharp quantum observables.  Clearly, the pairing of spinons
in the superconducting state is precisely the pairing that gives
rise to superconductivity.  However, while this connection is useful
for intuitive purposes, we feel that this state should probably not be
referred to as a spin liquid, and so we propose adding to the above
definition of a spin liquid the condition that large-scale gauge invariance 
(in the usual sense of superconductivity) should
also be an unbroken symmetry of the ground state.  The second example
is afforded by some quantum Hall liquid states of electrons with spin
\cite{balfrad}.  For instance, in a quantum Hall system consisting of
a Laughlin liquid\cite{laughliq} of strongly-paired opposite spin
electrons at filling factor $\nu=2$, it is easy to see that there
exist quasiparticles with spin 1/2, charge 0, and semionic
statistics\cite{tikof}.  This sort of state is a realization of the so-called
chiral spin liquid\cite{RL,wenzee}.


\begin{references}
\bibitem{tunnel} 
Note that single-particle tunnelling between the stripe and the environment 
is not relevant at low energies, since the constraint of
momentum conservation in the direction along the stripes and the
mismatch in Fermi wave vectors between the 1DEG and the environment
imply that it only has matix elements to high energy intermediate states;
this effect is further exaggerated by the presence of a spin gap.
\bibitem{diverge}  
In the 1DEG, the separation of spin and charge implies that all 
space-time collective susceptibilities are products of spin and charge
contributions. In the absence of a spin gap, the spin part of the
superconducting and the $2k_F$ CDW susceptibilities falls like
$\chi_s\sim [x^2+(v_st)^2]^{-K_s/2}$ with $K_s\approx 1$, otherwise, they  
approach a constant value proportional to the magnitude $\Delta_s$ of the spin
gap.  In the limit $\vec k$ and $\omega \rightarrow 0$, the temperature
dependence of the Fourier transforms of these susceptibilities behave as 
$\chi_{CDW}\sim T^{K_c + K_s-2}$ and $\chi_{sp}\sim T^{K_c^{-1} + K_s-2}$ 
in the absence of a spin gap, and $\chi_{CDW}\sim \Delta_s T^{K_c -2}$ and
$\chi_{sp}\sim \Delta_s T^{K_c^{-1} -2}$ in the presence of a spin gap.
Thus the opening of the spin gap enhances both susceptibilities at low 
temperatures by a factor of approximately $\Delta_s/T$ \cite{Boso1D}
\bibitem{expdef} Sometimes \cite{Boso1D} the notation  
($\theta_c$, $\theta_s$) is used for the parameters ($K_c$, $K_s$). 
The change in notation is introduced to avoid confusion with the boson 
phase variables $\theta_c$ and $\theta_s$.
\bibitem{freeze}  Because of the large environmental charge gap, fluctuations
of $\tilde\phi_c$ are effectively frozen, so the backscattering term
in Eq. (\ref{eq:He2}) is approximately given by simply the cosine of
the enivronmental spin field.
\bibitem{provided} 
Of course it is possible\cite{muttalib} to diagonalize the quadratic part of
the charge Hamiltonian for arbitrary $J_c$ and $V_c$, but we shall assume the
particular relation (\ref{eq:lambda}) because it simplifies the algebra, and
has no particular physical consequences. Moreover,
with this assumption, the coupling between the stripe and the
environment reduces to the usual forward scattering interaction
$\rho_1 \tilde \rho_2 + \rho_2 \tilde \rho_1$,
when $\tilde K_c = K_c =1$ and $\tilde v_c = v_c$.
\bibitem{dense} 
There is a fermionic correlation length $\xi_s = v_F/\Delta_s$
associated with the spin gap $\Delta_s$. When $\xi_s$ is much bigger than the characteristic
distance between impurity sites, $b = a/c \ll \xi_s$, the discrete character
of the impurity array can be ignored. Therefore, in this limit (high density),
we can also take the continuum limit with respect to the impurity lattice.
The spin Hamiltonian now takes the regular quantum sine-Gordon form
$\tilde H_s = H_0^s$, with $\beta = \sqrt {2 \pi}$.
\end{references}
\end{document}